\documentclass[%
 aip,
 amsmath,amssymb,
 reprint,
]{revtex4-1}

\usepackage{graphicx}
\usepackage[caption=false]{subfig}
\usepackage{dcolumn}
\usepackage{bm}

\usepackage[utf8]{inputenc}
\usepackage[T1]{fontenc}
\usepackage{mathptmx}

\usepackage{color}

\emergencystretch=\maxdimen
\begin{document}

\title{Particle Deceleration for Collective QED Signatures}
\author{A. Griffith}
\email{arbg@princeton.edu}
\author{K. Qu}
\author{N. J. Fisch}
\affiliation{Department of Astrophysical Sciences, Princeton University, Princeton, New Jersey 08540, USA}
\date{\today}

\begin{abstract}
Frequency upshifts have been proposed as a first experimental signature of collective effects in QED cascade generated electron-positron pair plasmas.
Since the high effective masses of generated pairs will reduce any frequency change, stopped pairs at minimal Lorentz factor in the lab frame were thought to be the dominant contribution to the the laser upshift.
However, we demonstrate that only considering stopped particles unduly neglects the contributions of particles re-accelerated in the laser propagation direction.
Re-accelerated particles should, on a per particle basis, affect the laser more strongly, and over a much longer timescale.
To maximize particle contributions to the laser upshift, we consider a Laguerre-Gaussian (LG) mode to better reflect generated pairs.
The LG mode doesn't have an advantage in particle deceleration and re-acceleration when compared against a Gaussian beam, but the LG mode can maintain particle contributions for a longer duration, allowing for more pair density accumulation.
Deceleration with a structured beam to keep pairs within the laser should create a larger upshift, thereby lowering the demands on the driving laser.
\end{abstract}

\maketitle

\section{Introduction}
\begin{figure}[tb]
    \includegraphics[trim={0 5 0 5},clip,width=\columnwidth, height=0.4\columnwidth]{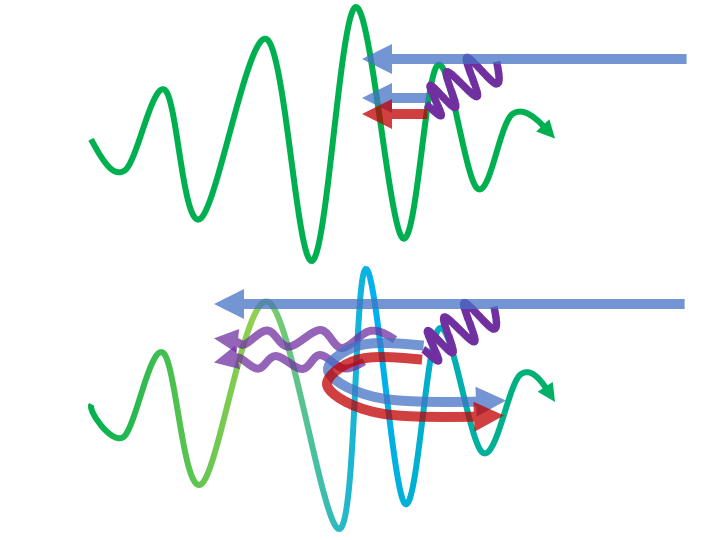}
    \caption{An electron beam (blue) generates high energy photons (purple) when colliding with a high intensity pulse (green). The photons may decay into electron (blue) and positron (red) pairs. These pairs decelerate through quantum synchrotron radiation, before being reflected by the Lorentz force. When the particles have low effective masses they upshift the driving laser pulse.}
    \label{fig:pair_generation_cartoon}
\end{figure}

It would be a great advance to investigate collective electron-positron pair plasma dynamics at experimental facilities.\cite{bell_possibility_2008,fedotov_limitations_2010, bulanov_schwinger_2010,zhu_dense_2016, yakimenko_prospect_2019}
Of particular interest is creating a region of high density pairs through a quantum electrodynamic (QED) cascade.\cite{di_piazza_extremely_2012}
A high density of electron-positron pairs will allow for experimental investigation of unique collective effects.\cite{nerush_laser_2011, grismayer_laser_2016, savin_energy_2019, edwards_strongly_2016,schluck_parametric_2017, gong_brilliant_2018, 
tiwary_particle_2021, huang_relativistic-induced_2021, li_production_2020}
However, distinguishing the unique aspects of collective pair plasma behavior at the short length and timescales of realizable experiments remains a challenge.
A possible method of probing collective effects was previously proposed for QED cascades originating from a laser-electron beam collision.\cite{qu_signature_2021, qu_collective_2021}
In a proposed experiment\cite{meuren_seminal_2020-1,meuren_mp3_2021, amaro_optimal_2021} the laser-electron beam collision produces high energy photons which then decay into electron positron pairs, diagramed in Fig.~\ref{fig:pair_generation_cartoon}.
At high enough laser and electron beam energies, the QED cascade can spawn a large number of pairs near the laser focus.
The resulting density of electrons and positrons can reach many multiples of the electron beam density, changing the plasma frequency.
Any change in the plasma density alters the dispersion relation for the passing driving laser, creating a corresponding frequency change.
In this manner, pair density and corresponding plasma frequency changes can induce a shift in frequency, similar to ionization.\cite{wood_measurement_1991, nishida_experimental_2012,qu_laser_2019}

Note that the single particle dynamics in a QED cascade differ greatly from electrons generated through ionization. 
The QED cascade occurs at much higher laser intensities and corresponding particle energies than ionization.
In the extreme regime required for the QED cascade, any generated pairs will be moving highly relativistically.
Highly relativistic particles will be much less responsive to the laser and their contribution to the plasma frequency will be much less.
This lessening will greatly reduce the impact of the particles on the laser, leading to a weak signature of collective QED effects.
To produce large upshifts, it is thus essential to slow down the pair particles.
Slowing the pair particles to reduce their Lorentz factor in the lab frame increases their contributions to the signature.\cite{qu_signature_2021}
Slowdown is achieved through a combination of the radiation reaction and the Lorentz force.
We provide a fuller treatment of deceleration than previous work to determine criteria on which particles can contribute and how deceleration can be optimized to increase the signature strength.
When particle deceleration is considered in more detail, we demonstrate that generated pairs have more potential to drive frequency upshifts.

Interestingly, it turns out that particle acceleration after reflection is favorable for creating discernable signatures.
In the rest frame of the re-accelerated particles, the frequency of the laser is downshifted and the critical density is lowered.
This Doppler downshift thus results in a larger plasma contribution and frequency shift.
The Doppler downshift in the laser overcomes increases of the particle Lorentz factor.
This results in larger frequency shifts than previously thought, accentuating the importance of particle reflection.
To improve particle acceleration we take inspiration from direct laser acceleration, and compare using a Gaussian laser field to using a Laguerre-Gaussian (LG) mode.
Direct laser acceleration theory suggests that the on-axis longitudinal laser field could be favorable for particle deceleration.
For the parameters we explored, the LG mode fails to improve particle deceleration and re-acceleration, but it does confine particles within the beam for longer durations.
Increased particle confinement should lead to increased density accumulation.
Combining re-acceleration and increased density accumulation in an LG mode should allow for stronger signatures.

This paper describes in detail how upshifts can be amplified by reflecting the generated electrons and positrons.
In Section~\ref{sec:weighting} we clarify the role of particle momentum in signal strength.
Deceleration will change not only the Lorentz factor, but also shifts the frame in which the critical density is defined, accentuating the role of the beam in signature generation.
In Section~\ref{sec:decel_guidelines} we formalize previously established guidelines on the laser $a_0$ required to stop a wide range of particle energies, widening the bounds on acceptable beam and laser configurations.
In Section~\ref{sec:case_studies} we numerically evaluate single particle dynamics and compare beam profiles to improve deceleration.
Section~\ref{sec:sum} provides a summary and discussion of the key results.

\section{Pair Contributions to Frequency Shifts}
\label{sec:weighting}
Electrons and positrons affect the laser frequency through both their spatial and momentum distribution.
Changes in the spatial distribution of electrons and positrons, namely the peak density, are initially driven by pair production in the QED cascade.
We review how the density of pairs drives frequency shifts, before elaborating on the dependency of frequency shifts on the momentum of the pairs to address some subtlety that was neglected in a previous publication.\cite{qu_signature_2021}

A sufficiently energetic laser and counter-propagating electron beam can produce through pair creation a large plasma density if the electric field exceeds the critical field, $E_c = m^2c^3/(\hbar e)$, in the rest frame of an electron.
The field strength relative to the critical field in the electron rest frame is characterized by the quantum nonlinear factor
\begin{equation}
    \chi = \frac{\gamma}{E_c}
    \sqrt{\left| \mathbf{E} + \mathbf{v}\times\mathbf{B}\right|^2 - (\mathbf{v}\cdot \mathbf{E})^2} \label{eqn:chi},
\end{equation}
for electron Lorentz factor $\gamma$ and velocity $\mathbf{v}$ and laser electric field $\mathbf{E}$ and magnetic field $\mathbf{B}$.
When $\chi \gg 1$ initially, a high intensity laser and an energetic and high density electron beam can produce an electron and positron plasma of more than an order of magnitude greater than the initial electron density.\cite{qu_signature_2021}

Creation of pair plasma in the laser field changes the disperison relation and upshifts the laser frequency.
For low plasma density, the frequency upshift is $(\Delta \omega_p)^2/(2\omega_0)$, where $\Delta \omega_p$ is the change in plasma frequency and $\omega_0$ is the laser frequency.
The plasma frequency accounting for the relavistic particle mass increase is
\begin{equation}
    \omega_p^2 = \frac{4\pi n_p e^2}{\gamma m},
    \label{eqn:omega_p_mod} 
\end{equation}
for pair density $n_p = n_{e^+} + n_{e^-}$, charge $e$, pair Lorentz factor $\gamma$, and pair mass $m$.
The density changes orders of magnitude, but the laser frequency, $\omega$, is changed less significantly.
With spatial and temporal dependence, the local frequency shift $\Delta\omega$ may be expressed\cite{qu_signature_2021} as
\begin{equation}
\frac{\Delta \omega(x,t)}{\omega_0} = \int_{t_0}^tdt' \partial_T \frac{n_p(X,T)}{n_c(X,T)\gamma(X,T)}\Big{|}_{X=x - c(t' - t)}^{T=t'},
\label{eqn:freq_shift}
\end{equation}
which is written in terms of ratio of the plasma density to the laser critical density $n_c = m \omega^2/(4\pi e^2)$.
To increase $\Delta \omega$, previous work\cite{qu_signature_2021,qu_collective_2021} focussed on maximizing $n_p$ and minimizing $\gamma$ to create a discernable signature.
To further magnify frequency shifts we elaborate here on previous work, and consider changes in the critical density, $n_c$, as a way to best amplify the frequency shift.

The most straightforward change one might consider is using a lower frequency driving or secondary probe beam to lower $\omega$ and in turn $n_c$.
Using a lower frequency laser is stymied by the tight focussing required to achieve a large $\chi$.
A maximally focussed lower frequency driving laser will smear out pairs over a larger volume, lowering $n_p$ and thus $\Delta \omega$.
Another alternative would be to use a lower frequency probe beam combined with a tightly focussed higher frequency drive beam.
However, in this configuration, the secondary probe will interact with a plasma volume smaller than the probe laser wavelength.
When the plasma volume is small compared to the wavelength, the scale separation required for the desired upshift effect is no longer valid.
We cannot use either of these approaches to lower the critical density and increase the signature.
Hence, to magnify $\Delta \omega$ by considering $n_c$, we work with the primary beam.

The critical density can change as the laser frequency the pairs experience is Doppler shifted.
When the pairs are stopped in the lab frame, the pairs oscillate at $\omega_0$, but if the pairs are moving parallel to the laser phase velocity, the laser oscillation period of the particles changes.
To account for the changing laser frequency as experienced by the pairs, the critical density is calculated in the frame where the plasma has no flow in the direction of the laser.
Pairs co-propagating or counter-propagating with the laser decrease or increase the critical density, respectively.
If the pairs co-propagate with the laser, the negative Doppler shift of laser frequency can decrease the critical density, resulting in a higher laser frequency shift.
Accounting for this frame change only the critical density in Eq.~\eqref{eqn:freq_shift} changes, as the term $n_p/\gamma$ is Lorentz invariant.
When the changing critical density is taken into account we can estimate how much each additional particle may contribute to frequency shifts based on the momentum of the particle
\begin{equation}
    \frac{\Delta \omega}{\Delta n_p} \propto \frac{1+\beta_z}{\gamma(1-\beta_z)},
    \label{eqn:weighting}
\end{equation}
for laser propagation direction $z$, pair Lorentz factor $\gamma$, and particle velocity $\beta_z = v_z/c$.
Particle contributions to $\Delta \omega$ will be suppressed or magnified by both the particle Lorentz factor and the laser Doppler shift.
To maximize any observable frequency change we aim to increase Eq.~\eqref{eqn:weighting} for any generated pairs.

This subtly changes our aims from previous work, which focussed on reducing only $\gamma$.
Equation~\eqref{eqn:weighting} more strongly suppresses counter-propagating particles of equivalent $\gamma$, and increases the impact of co-propagating particles.
Not only will pairs at the point of reflection where $\gamma$ is minimized contribute, but particles re-accelerated such that increases in $\gamma$ are dominated by decreases in $(1-\beta_z)/(1+\beta_z)$ can contribute even more strongly.
Considering re-acceleration increases the duration for which particles are relevant for frequency shifts, and affects the aims of particle deceleration and reflection.

\section{Particle Deceleration and Reflection}
\label{sec:decel_guidelines}
In a QED cascade, pair frequency-shift contributions will initially be small.
Pairs will primarily be generated with highly relavistic momenta anti-parallel to the laser propagation direction with $-\gamma\beta_z \gg 1$, making Eq.~\eqref{eqn:weighting} negligible.
However, the laser provides an opposing force which can reverse the particle momentum and increase pair frequency contributions.
We now detail how the laser can reduce $\gamma$ and reverse $\beta_z$ to create a significant frequency shift.

Changes in $\gamma$ and $\beta_z$ occur through a combination of the radiation reaction and the Lorentz force.
Initially, when $\gamma$ is large, the dominant effect is the radiation reaction.
For large $\gamma$ and $\beta_z \rightarrow -1$, $\chi > 1$ and the pairs stochastically emit high energy photons through quantum synchrotron radiation.
The quantum radiation reaction will provide the dominant reduction in $\gamma$, but will not significantly change $\beta_z$ as for $\gamma \gg 1$ orders of magnitude changes in $\gamma$ correspond to minimal changes in $\beta_z$.
Before $\beta_z$ sufficiently differs from $-1$, the Lorentz factor of any pairs will have been reduced to such a degree that $\chi < 1$ and the forcing on the particle from quantum synchrotron radiation has greatly weakened.

The sign change of $\beta_z$, corresponding to reflection of the particle, is instead driven by a longitudinal Lorentz force.
This can only occur if the radiation reaction has successfully decelerated particles to a low enough $\gamma$ such that the Lorentz force is not heavily suppressed by the high effective particle mass.
For the Lorentz force to work in concert with the radiation reaction, the radiation reaction must be able to successfully decelerate particles to a low enough $\gamma$ such that the Lorentz force can reflect generated pairs.
Starting from the Lorentz force, we determine a maximum reflectable pair energy, and use this condition to determine the laser intensity required to decelerate particles down to this scale.

\subsection{Reflection}
\label{sec:lorentz}
Particles will be reflected through electrically driven transverse oscillations creating a longitudinal $\mathbf{v}\times\mathbf{B}$ response.
A transverse electric field is the focus of our analysis, as a directly longitudinal electric field cannot simultaneously provide a large quantum nonlinear factor due to the $\mathbf{v}\cdot\mathbf{E}$ term in Eq.~\eqref{eqn:chi}.
A large quantum nonlinear factor is required for both pair production and decelerating particles through quantum synchrotron radiation.
Both factors are required for the proposed experimental test of collective QED effects, and so we rely on $\mathbf{v}\times\mathbf{B}$ forcing to provide the longitudinal work.

We next derive a simple estimate of particle contributions in a transversely uniform plane with vector potential $\mathbf{A}$.
The vector potential, neglecting the minimal plasma contribution, can be written as $\mathbf{A} = A_0 g(\phi)\cos(\phi)\mathbf{\hat{x}}$ for $\phi = \omega (t - z/c)$ with slowly varying envelope $d_\phi g(\phi) \ll g$.
For an electromagnetic wave which purely depends on $\phi$ there is the symmetry $t \rightarrow t + \lambda,\ z \rightarrow z + c\lambda$ with the corresponding conserved quantity for particle motion
\begin{equation}
    E - cp_z = mc^2 \gamma(1-\beta_z).
    \label{eqn:consv}
\end{equation}
Equation~\eqref{eqn:consv} is important, as it bounds frequency contributions before we consider the single particle dynamics.
As a consequence of Eq.~\eqref{eqn:consv}, through reflecting the particle and changing $\beta_z$ from negative to positive, the particle energy must increase in the lab frame.
Previously, this was believed to suppress the signal.
However, this change in energy doesn't dampen particle contributions as the denominator of Eq.~\eqref{eqn:freq_shift} remains constant, and the numerator can greatly increase as a particle initially with $\beta_z \rightarrow -1$ is reflected.

We continue by writing out particle dynamics as a function of $\phi$, following Hartemman et al.\cite{hartemann_nonlinear_1995}
Dimensionless equations of motion for positrons and electrons can be written as
\begin{align}
    d_\phi (\gamma \beta_x) &= \pm a_0 g(\phi)\sin(\phi),  \\
    d_\phi (\gamma \beta_z) (1-\beta_z)&= \pm a_0\beta_x g(\phi)\sin\phi,
\end{align}
for normalized vector potential $a_0 = \frac{eA_0}{mc^2}$.
Through integrating and taking advantage of Eq.~\eqref{eqn:consv},  the pair energy $\gamma$  can be written as a function of $\phi$
\begin{equation}
    \gamma(\phi) = \gamma_0 \left [1 + a_0^2 \frac{1+\beta_z(\phi_0)}{2}\left(\int_{\phi_0}^\phi g(\phi)\sin(\phi)d\phi\right)^2  \right],
    \label{eqn:gammaphi}
\end{equation}
where $\phi_0$ is the phase at which the last photon recoil occurs.

The shape of the laser envelope, $g(\phi)$, should not strongly affect the particle dynamics which are relevant for creating QED cascade signatures.
The laser envelope influences the particle Lorentz factor through the integral in Eq.~\eqref{eqn:gammaphi} over many laser cycles.
This can be interpreted as the ponderomotive force, which is determined by the gradient of the laser amplitude.
The ponderomotive force is, however, negligible in the region where the pairs are created and slowed through radiation reaction.
This is because the QED process happens in the region where $g(\phi)$ is maximized such that $\chi \gg 1$.
Where $g(\phi)$ is peaked, the gradient in the laser field strength is zero to first order and the longitudinal ponderomotive force is weak.
Near the focus the integral will only oscillate around the value of $g$ and not accumulate changes of  $g$ to influence $\gamma$. 

Within single laser cycles, the average drift of the particles in the laser will increase the  average  $\gamma$.
For $\phi$ varying over a single cycle, forcing comes from changes in the sign of $\sin(\phi)$, corresponding to single cycle laser acceleration.
Under the slowly varying envelope approximation, the particle energy will change in proportion to $(g(\phi)\cos\phi - g(\phi_0)\cos\phi_0)^2$, which near the laser peak may be further approximated as $g(\phi_0)^2(\cos\phi - \cos\phi_0)^2$.
The particle momentum will oscillate, but the average of this quantity will be proportional to $1/2+\cos^2\phi_0$. 
Depending on the initial phase $\phi_0$ there will be some average increase in $\gamma$ and a  corresponding drift while the pair particle is in the laser.
The initial phase will most likely occur near a peak of $|\sin\phi|$, as both pair production and quantum synchrotron radiation scale increasingly with field strength. 
Extrema of $\sin\phi$ correspond to roots of $\cos\phi$, so the magnitude of the integral will be minimized and the phase average of Eq.~\eqref{eqn:gammaphi} will be close to $1/2$ for most particles.
Importantly, the particle drift will cease when the laser passes, but temporary changes of particle momenta are sufficient for generating a large signature, unlike in particle acceleration which aims to maximize the final momentum of the particles.

Oscillatory changes in particle momentum should be sufficient to increase particle contributions to the frequency shift.
For initially highly relativistic pairs particles with $\beta_z(\phi_0)\rightarrow -1$ the contribution to the frequency shift can be approximated through using Eq.~\eqref{eqn:consv} as
\begin{equation}
    \frac{1+\beta_z}{\gamma(1-\beta_z)} \approx \gamma_0^{-1} - \gamma(\phi)^{-1}.
    \label{eqn:freq_approx}
\end{equation}
The Lorentz factor will oscillate, but when the particle is being driven by the Lorentz force the average value of $\gamma(\phi)$ will exceed $\gamma_0$.
If the average value of $\gamma(\phi)$ is significantly larger than $\gamma_0$ while a particle oscillates in the wave then particle effects on the laser frequency are magnified.

For $\gamma(\phi)$ to be significantly larger than $\gamma_0$ the laser strength $a_0$ must overcome the suppression from $1+\beta_z(\phi_0)$ in Eq.~\eqref{eqn:gammaphi}.
This corresponds to the condition that 
\begin{equation}
    a_0^2 \frac{1+\beta_z(\phi_0)}{2} > 1,
\end{equation}
which implies that for particles to contribute significantly more that $a_0(1- a_0^{-2})^{-1/2} > 2\gamma_0$.
For $a_0 \gg 1$, as is the case here, this maps to the intuitive condition that Lorentz force will only be relevant when particles are decelerated down to energies on the order of the laser potential.

Physically, the requirements on $a_0$ may be understood as a requirement that particles must be reflected at some point within the laser to contribute strongly.
If quantum synchrotron radiation ceases at phase $\phi_0$, leaving the particle with velocity $\beta_z(\phi_0) = -\beta_0$, then for a particle to be reflected at some later phase $\phi_r$ requires that
\begin{equation}
    a_0^2\left(\int_{\phi_0}^{\phi_r} g(\phi)\sin(\phi)d\phi\right)^2 = \frac{2\beta_0}{1-\beta_0}.
    \label{eqn:maxbeta}
\end{equation}
Given that $g(\phi)$ is a slowly varying envelope scaled to unity the integrand is at most $\sim 2$ and the maximum initial particle Lorentz factor $\gamma_{\textrm{lim}}$ which can be stopped is
\begin{equation}
    \gamma_{\textrm{lim}} \sim \frac{1+2a_0^2}{\sqrt{1+4a_0^2}}.
     \label{eqn:energy_lim}
\end{equation}
For $a_0 \gg 1$ this corresponds to the same scaling on the laser intensity that $\gamma \sim a_0$, only differing by a factor of $2$ as we have assumed that the phase is chosen to maximize the integral quantity.
Strong contributions of pair particles require for changes in $\beta_z$ to be significant enough such that $1 + \beta_z$ is no longer negligible.
An even stronger laser and corresponding re-acceleration may increase this quantity further up to a factor of $2$ after reflection, but this is dwarfed by earlier changes of orders of magnitude.

The scale the the Lorentz force, where $\gamma_0 \sim a_0$, sets the requirements of the damping that must be provided by the radiation reaction.
Successive recoils from emitted photons must drive particles down from initial energies which greatly exceed $a_0$ to this scale for both factors to work in tandem to produce large signatures.

\subsection{Lorentz Force - Quantum Synchrotron Radiation Interplay}
\label{sec:quasynchrotron}
Initially high pair particle energies are lowered in the high field regime, where $\chi \gg 1$, through quantum synchrotron emission.\cite{blackburn_radiation_2020}
Particles stochastically emit high energy photons in the strong laser field, resulting in a recoil opposite their velocity.
The electrons and positrons emit photons with probability per unit time $dW$ with photon energy, $\mathcal{E}$, over a distribution
\begin{equation}
    \begin{split}
     \frac{dW}{d\mathcal{E}} = \frac{\alpha\lambda}{\lambda_c\sqrt{3}\pi \gamma^2 (1+u)} \left( \vphantom{\int_1^2} (1 + (1 + u)^2)K_{2/3}(\xi)  \right.\\
     -\left. (1 + u)\int_0^\infty K_{1/3}(y + \xi) dy\right)
    \end{split}
    \label{eqn:prob_dist}
\end{equation}
for fine structure constant $\alpha$, Compton wavelength $\lambda_c$, $u = \frac{\mathcal{E}}{\gamma-\mathcal{E}}$, $\xi = 2u / 3\chi$, and $K_\nu$ being the $\nu$'th Bessel function of the second kind.\cite{tamburini_laser-pulse-shape_2017}
Unlike the Lorentz force, high anti-parallel particle momentum doesn't  suppress the work done through this effect, but increases it due to the asymptotic exponential dependence of $K_\nu$ on $\xi$.

The quantum synchrotron radiation is the dominant form of deceleration initially when $-p_z \gg m_e c$.
As the magnitude of the momentum drops, the expected frequency of photon emission decreases.
If the laser and particle energy scales are well matched, the Lorentz force can provide the remaining necessary work right after photon emission ceases.
The interplay between these two mechanisms enters here, where the aim is to decelerate particles in ways that do not enter in the case of traditional vacuum laser acceleration.

We set a simple scaling for which particles can be completely stopped based on Sec.~\ref{sec:lorentz}.
As a rough guideline, there should be no gap in the range of particle energies at which the two mechanisms act.
This requires that photons are still being emitted at the maximum energy at which the Lorentz force can act effectively.
Based on Eq.~\eqref{eqn:energy_lim} the particle can be stopped after the radiation reaction is weak, approximately if $\gamma = \gamma_{\textrm{lim}}\sim a_0$.
For this transition to occur, the radiation reaction should provide a last decelerating impulse at a peak of the oscillating electric field.
The last photon emission must thus occur when the particle is at $\gamma \sim a_0$.
Photon emission is strongly dependent on $\chi$, and should remain significant while $\chi > 0.1$.\cite{qu_signature_2021}
In a plane wave geometry an approximate lower bound for photon emission is thus $0.1 = \chi \sim 2 \gamma a_0\lambda_c/\lambda$.
Combining both scales, reflecting high energy particles requires a minimum field $a_{\textrm{cutoff}}$, that must satisfy $a_{\textrm{cutoff}}^2 >\frac{1}{20} \lambda/\lambda_c$.
For an electromagnetic wave with $\lambda = 0.8\mu$m this implies a minimum $a_0 \sim 130$ for a large number of particles to be stopped.
This is an imprecise estimate, as the rate and magnitude of photon emission will vary significantly depending on the pair particle energy as the particle slows.

\begin{figure}[t]
    \includegraphics[trim={10 0 0 5},clip,width=\columnwidth]{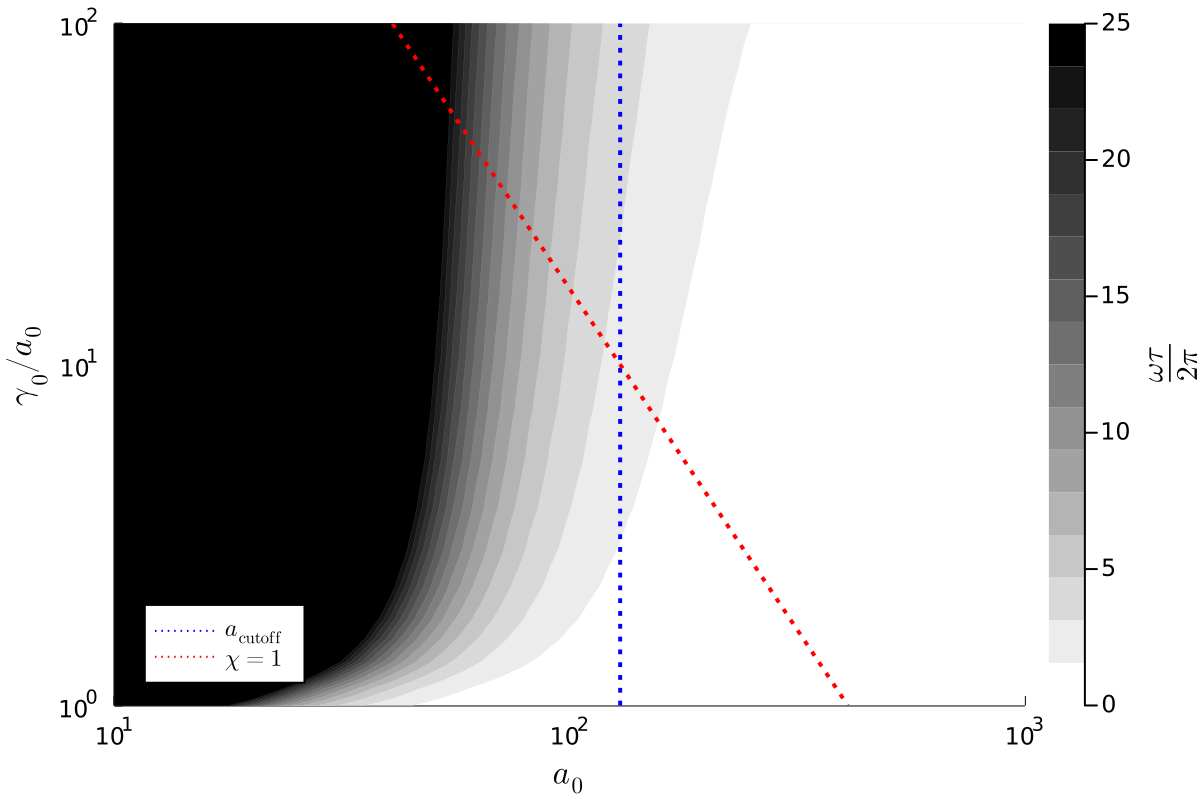}
    \caption{Decay time $\tau$ shown in the number of laser cycles for particle deceleration.
    For each $\gamma_0$ and $a_0$, $\tau$ is evaluated according to Eq.~\eqref{eqn:ode_gamma} assuming a constant field.
    The $\chi = 1$ boundary is also plotted for reference, along with $a_{\textrm{cutoff}}$.
    Viable experiments must generate particles well above the $\chi = 1$ bound, and with a low enough $\tau$ for generated particles to ensure the changing density contributes to $\Delta \omega$.}
    \label{fig:time_map}  
\end{figure}

To improve upon this cutoff, we suppose the quantum synchrotron radiation decelerates pair particles with the mean recoil across the full range of energy.
For initial particle energy $\gamma_0$ and laser field strength $a_0$,
we estimate how long the radiation reaction takes to drop the particle energy down to the scaling set by Eq.~\eqref{eqn:energy_lim}.
Assuming that the particle experiences the average forcing from the radiation reaction, the energy evolves according to the first moment of Eq.~\eqref{eqn:prob_dist}.
Integrating the photon emission rate, weighted by the photon energy across the spectrum, results in a decay of pair particle energy
\begin{equation}
    \frac{d\gamma}{dt} = -\int_0^\gamma \mathcal{E}\frac{dW}{d\mathcal{E}} d\mathcal{E}.
    \label{eqn:decay}
\end{equation}
If the particle starts from energy $\gamma_0$ and may be stopped when the energy reaches Eq.~\eqref{eqn:energy_lim}, then by integrating the inverse of Eq.~\eqref{eqn:decay} we can estimate a time $\tau$ for the energy decay to occur 
\begin{equation}
   \tau = \int_{\gamma_{\textrm{lim}}}^{\gamma_0} \left(\int_0^\gamma \mathcal{E}\frac{dW}{d\mathcal{E}} d\mathcal{E}\right)^{-1}d\gamma,
    \label{eqn:ode_gamma}
\end{equation}
where we have assumed a constant field strength of $a_0$.
For a range of the values of $a_0$ and $\gamma_0/a_0$, $\tau$ is plotted in Fig.~\ref{fig:time_map}.
The time scale is primarily determined by the strength of the laser $a_0$, with the contour levels of $\tau$ running almost parallel to the $\gamma_0/a_0$ axis.
The $a_{\textrm{cutoff}}$ pessimistically overestimates the $a_0$ required to stop particles in a feasible number of laser cycles.
Lasers with $a_0 \geq 100$ should stop a wide range of particles in relatively few laser cycles, resulting in a detectable signature.
At low $a_0$ the number of laser cycles, $\omega \tau / 2\pi$, dramatically exceeds the duration of any realizable laser.
If deceleration is the limiting constraint on generating a large signature, then accepting multi-cycle deceleration suggests that the constraint $a_0 \geq a_{\textrm{cutoff}}$ might be relaxed and laser power could be reduced by up to forty percent by taking $a_0$ from $130$ to $100$.

There may be an advantage to using this lower laser power for signature generation.
The laser $a_0$ will determine the initial $\gamma_0$ from which the pair is re-accelerated.
A lower $\gamma_0$ is favorable, as it increases the estimated frequency contribution in Eq.~\eqref{eqn:freq_approx}.
However, a higher $a_0$ will result in a lower time average of $\gamma(\phi)$ which from the same expression can be seen to increase pair frequency shift contributions.
A reduced baseline particle energy is balanced against less laser power being available for re-acceleration.

\section{Single Particle Case Studies}
\label{sec:case_studies}
To improve upon the simple but analytically tractable electromagnetic plane wave we considered in Sec.~\ref{sec:decel_guidelines}, we evolved single particle dynamics numerically in paraxial laser modes.
This includes the two dimensional variation which will occur in any realizable experiment as pulses will be tightly focussed to achieve the high intensities for pair generation.
Beyond extending our previous analytic work, numerical single particle evolution gives a clearer picture of the dynamics present in previous PIC results and offers the opportunity to further optimize signature generation.
Based on known results in particle acceleration we consider using an LG mode to improve performance in comparison to a Gaussian beam.
In comparing focussed beams it is clear that not only maximizing particle deceleration and re-acceleration, but keeping the particles within the intense region of the beam will be key to creating detectable signatures.

\subsection{Methodology}
Our numerical work is restricted to single particles in paraxial laser fields.
This will not capture the desired collective effect which requires the contributions of many particles.
But, if collective effects are weak, evolving test particles individually should be reasonably accurate.
Furthermore, previous work,\cite{poder_experimental_2018} comparing PIC against collective effect free particle evolution while examining the radiation reaction in a similar setup suggests the average dynamics will be highly similar.

\begin{figure*}[htb]
    \subfloat[\label{fig:gauss}]{
        \begin{tabular}[b]{c}
            \includegraphics[trim={90 10 90 55},clip,width=0.49\textwidth]{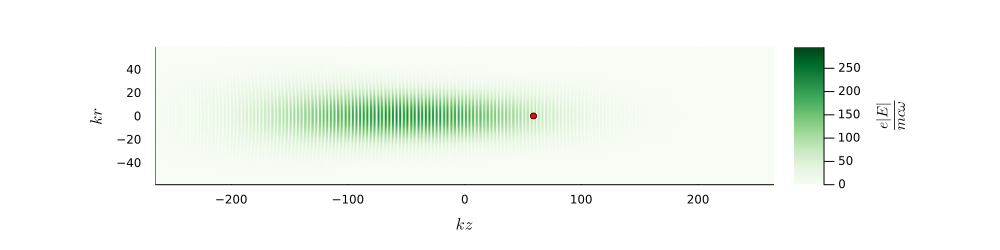}\\
            \includegraphics[trim={90 10 90 55},clip,width=0.49\textwidth]{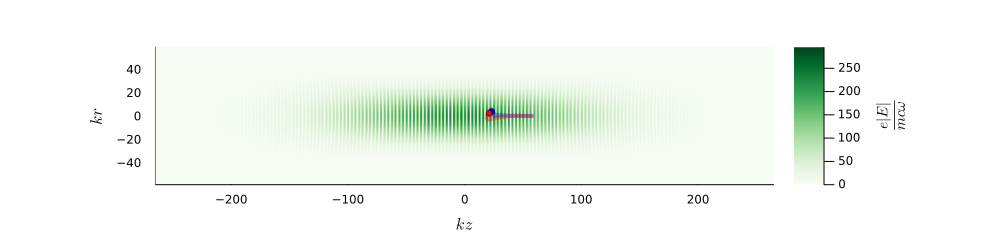}\\
            \includegraphics[trim={90 10 90 55},clip,width=0.49\textwidth]{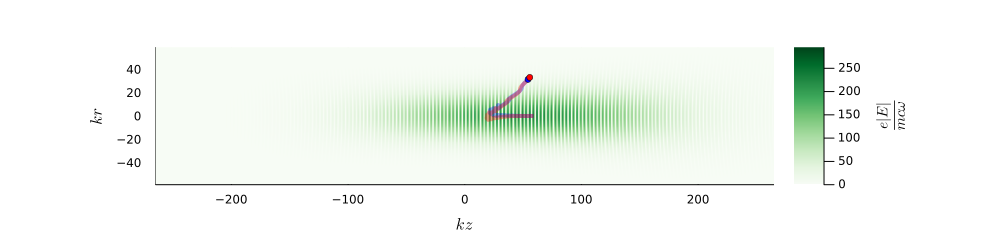}
        \end{tabular}
    }
    \subfloat[\label{fig:laguerre}]{
        \begin{tabular}[b]{c}
            \includegraphics[trim={90 10 90 55},clip,width=0.49\textwidth]{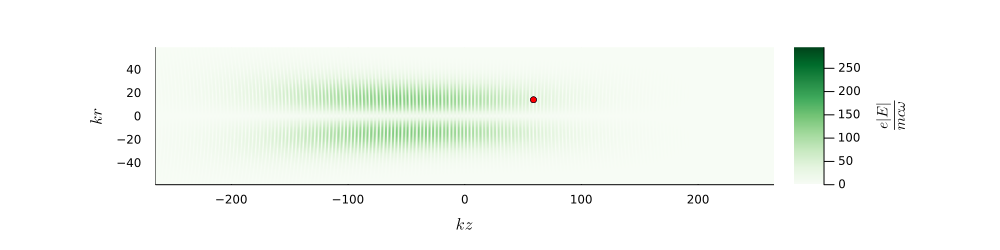}\\
            \includegraphics[trim={90 10 90 55},clip,width=0.49\textwidth]{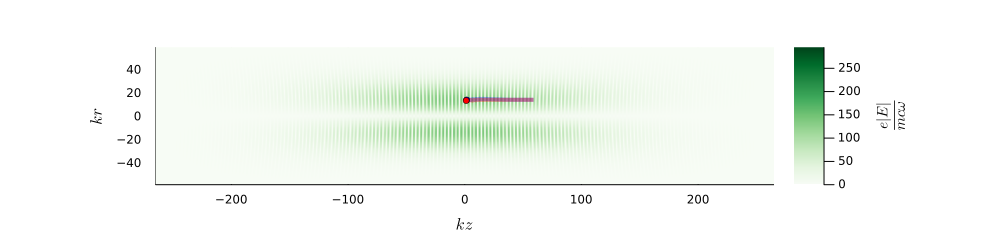}\\
            \includegraphics[trim={90 10 90 55},clip,width=0.49\textwidth]{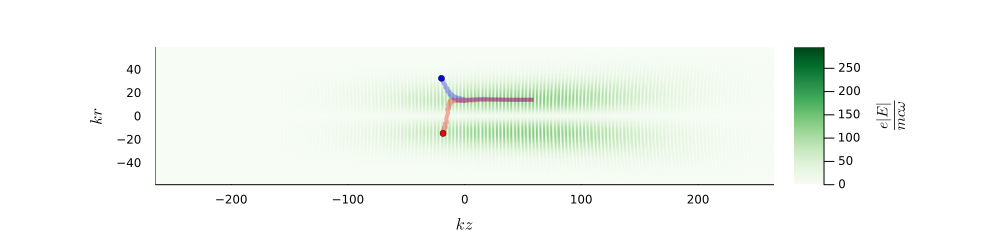}
        \end{tabular}
    }
    \caption{Trajectories of an electron (red) and positron (blue) in Gaussian (Fig.~\ref{fig:gauss}) and LG$_{l=1}$ (Fig.~\ref{fig:laguerre}) laser fields.
    Particles are initialized at the same $z$ position in the rising edge of the pulse, with the radial position in line with the maximum of the field strength in the focal plane.
    Particles eventually drift outside the laser beam, however, the LG mode can maintain co-propagation for a longer duration as demonstrated by the second snapshot where both the electron and positron are still within the LG beam, but have been scattered by the Gaussian beam.}
    \label{fig:frames}
\end{figure*}

\begin{figure*}[htb]
    \subfloat[\label{fig:dw_gauss}]{
        \includegraphics[trim={50 50 50 50},clip,width=0.49\textwidth]{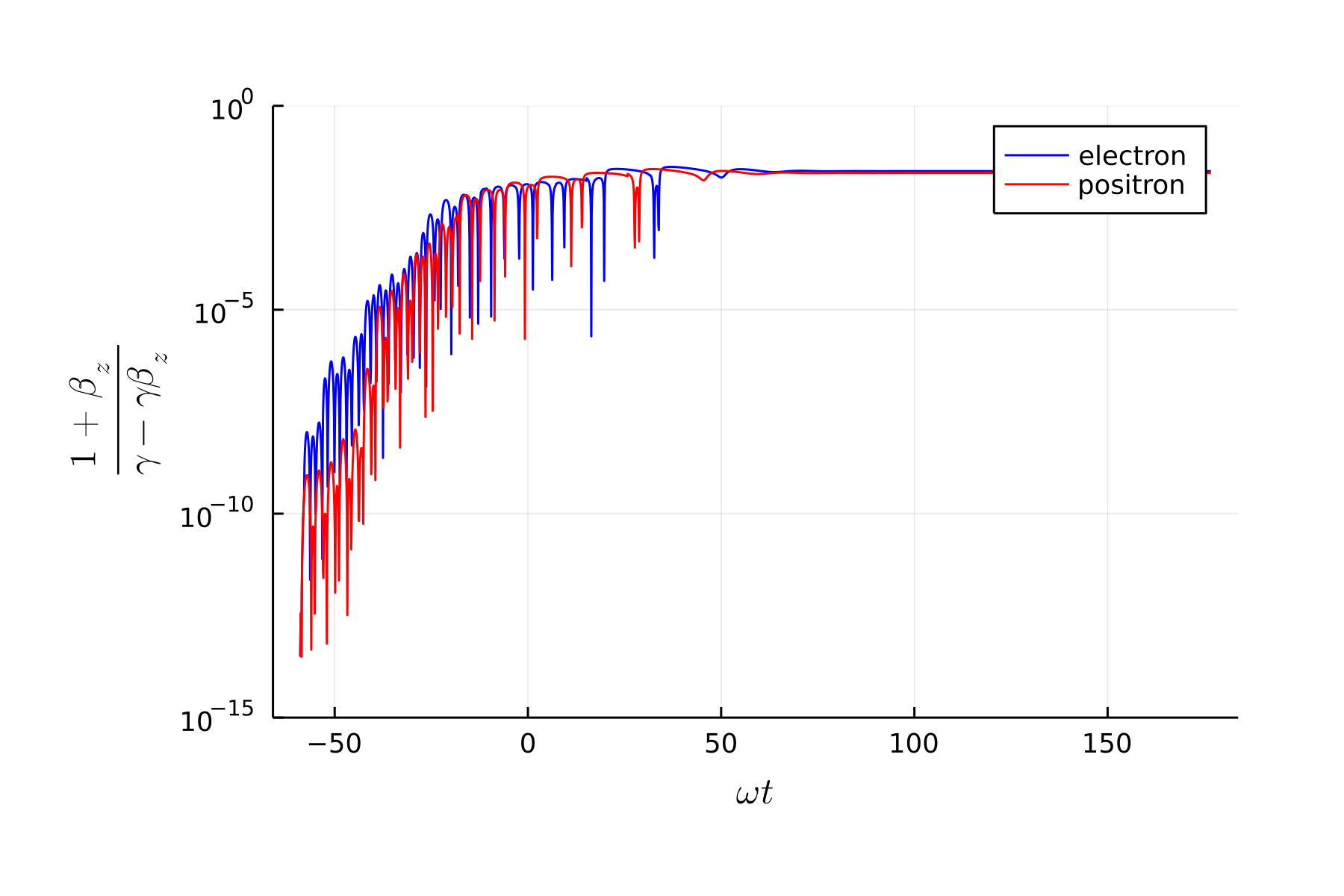}
    }
    \hfill
    \subfloat[\label{fig:dw_laguerre}]{
        \includegraphics[trim={50 50 50 50},clip,width=0.49\textwidth]{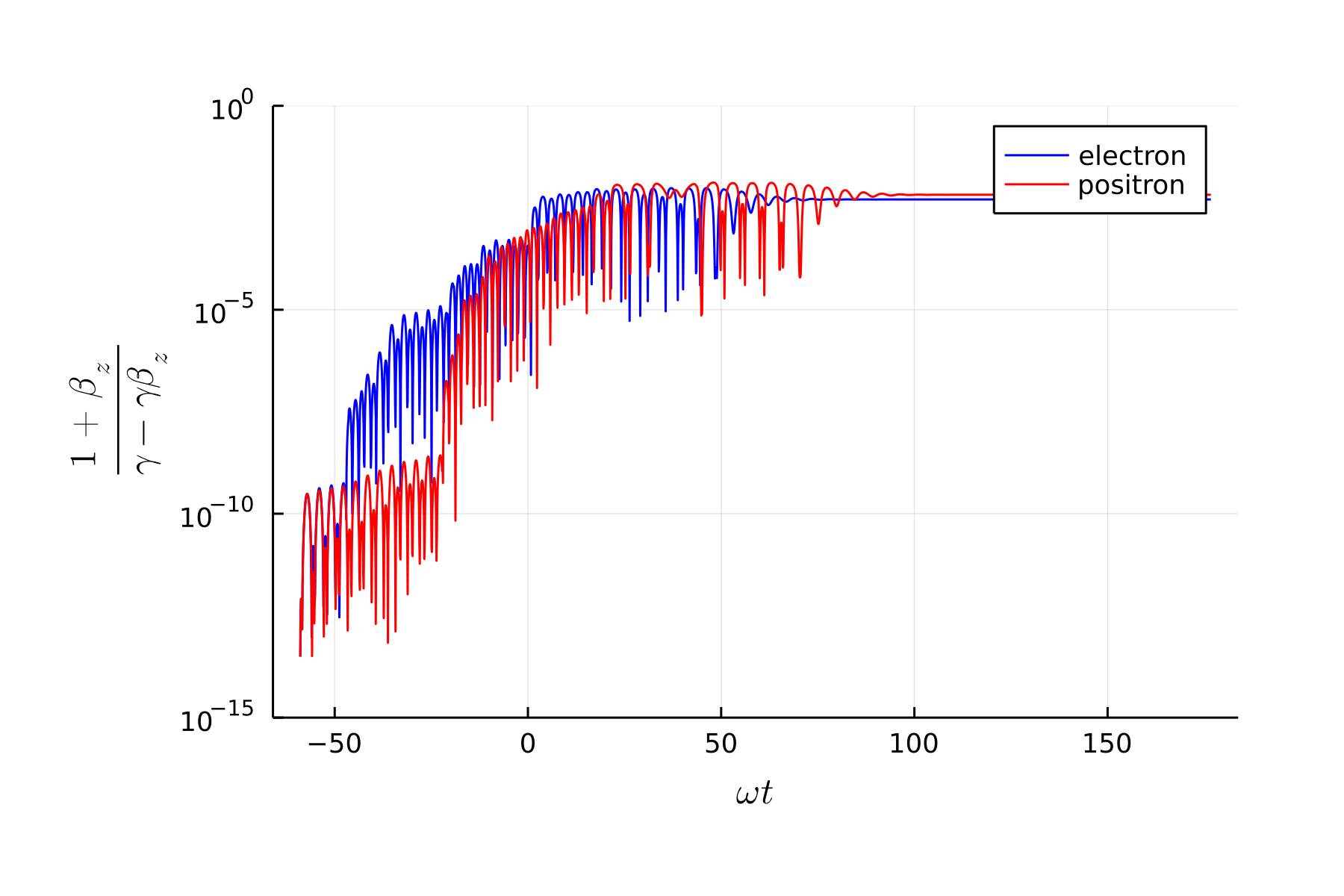}
    }
    \caption{Electron (red) and positron (blue) particle contribution weightings (Eq.~\eqref{eqn:weighting}) for Gaussian (Fig.~\ref{fig:dw_gauss}) and LG (Fig.~\ref{fig:dw_laguerre}) modes corresponding to trajectories from Fig.~\ref{fig:gauss} and Fig.~\ref{fig:laguerre} respectively.
    The Gaussian beam with higher peak intensity re-accelerates the particles to a higher $\beta_z$, however it cannot maintain the particles within the beam for long, reducing the number of oscillations each particle experiences within the beam compared to the LG mode. Discrete jumps in momentum caused by photon emission occur at early times, giving way to later slower timescale oscillations.}
    \label{fig:dw}
\end{figure*}

Particles are evolved using a standard differential equation solver\cite{rackauckas_differentialequationsjl_2017} in a prescribed laser field.
The laser field is one of various electromagnetic Laguerre-Gaussian modes with a temporal Gaussian envelope.
Particles are initialized with a purely negative $z$ momentum and are primarily evolved according to the Lorentz force.
Quantum synchrotron radiation is modeled through randomized reductions in the particle energy.
Both the rate and distribution of radiated photon energy are handled stochastically in accordance to Eq.~\eqref{eqn:prob_dist} following the procedure outlined in the supplementary material of Tamburini et al.\cite{tamburini_laser-pulse-shape_2017}

\subsection{Particle Deceleration with a Gaussian Laser Field}
We use particle evolution in a Gaussian beam to compare against both the analytical estimates in Sec.~\ref{sec:decel_guidelines} and the LG mode in Sec.~\ref{sec:subLG}. 
A positron and electron are initialized, shown in the first panel of Fig.~\ref{fig:gauss}, with $10$ GeV in the rising edge of a $50$ fs pulse with a peak intensity of $6\times10^{22}$ W/cm$^2$ focussed to a beamwidth of $5\ \mu$m.
While the particles have high counter-propagating momentum quantum synchrotron radiation dominates, but photon emissions falls off as the Lorentz factor of the particles drops to around the laser $a_0 = 300$ and the particles change direction, as shown in the second panel of Fig.~\ref{fig:gauss}.
The particles then oscillate with an increasing period and start to move in the positive $z$ direction over many cycles as $\mathbf{v} \times \mathbf{B}$ provides longitudinal work.
Eventually the particles exit the laser with positive $z$ momentum, as shown in the last panel of Fig.~\ref{fig:gauss}.

In this simulation particle re-acceleration leads to higher expected particle contributions to the frequency shift.
The reduction in particle energy along with the increase in particle $\beta_z$ implies that the contribution of the particles to the frequency change given by Eq.~\eqref{eqn:weighting} rapidly increases.
The increase in oscillation period in tandem with the increasing particle contribution factor can be seen in Fig.~\ref{fig:dw_gauss}.
Contributions remain at level where they are still suppressed when compared to particles at rest, where $\gamma = 1$ and $\beta_z = 0$, but the increase over time is highly significant.

The benefit of re-acceleration is limited as increases in Fig.~\ref{fig:dw_gauss} saturate.
This saturation may be understood as being due to the fact that, in the later stages, when $\gamma(1- \beta_z)$ should be conserved, the particle contribution changes only due to shifts of $1+\beta_z$, which can only increase particle contributions by up to a factor of $2$ as the particle is re-accelerated compared to the point of reflection.
The particle eventually exits the beam, and is then no longer able to influence the laser.
The increase in particle contributions to the frequency shift in Fig.~\ref{fig:dw_gauss} after $\omega t = 0$ demonstrates that signatures should be increased by re-acceleration, but contributions are terminated relatively early.

Laser driven re-acceleration can improve the strength of the signature, but the Gaussian beam reduces the impact of each particle due to transverse scattering.
The ponderomotive potential of the Gaussian field is peaked on axis so particles tend to be pushed out of the pulse.
When the particles radially exit they no longer interact with the beam and are no longer relevant to signature generation.
Moreover, for a very tightly focussed Gaussian beam particles may be ejected before experiencing significant re-acceleration.
Transverse scattering will also reduce the density of particles as they are spread out over a larger volume.
All of these factors stunt particle contributions to the signature, and could prove to be greatly limiting.
Completely uncontrolled particle dispersal may reduce and even reverse any gains from particle deceleration on the frequency shift.

\subsection{Comparing Particle Deceleration with a Laguerre-Gaussian Laser Field}
\label{sec:subLG}
As an alternative we considered using an LG $l=1$ mode, as an example of a structured beam to enhance particle reflection and the collective effect signature.
LG modes are well studied for both particle acceleration\cite{varin_acceleration_2002,karmakar_collimated_2007, fortin_direct-field_2009, esarey_physics_2009} and electron-positron pair generation.\cite{elkina_qed_2011,tamburini_laser-pulse-shape_2017,mercuri-baron_impact_2021-1}
LG modes have useful structure, and radially polarized beams can provide a strong on axis longitudinal field.
The longitudinal field is favorable for direct laser acceleration.\cite{arefiev_novel_2015}
This motivation drove our consideration of LG modes.

Our simulations show that deceleration and re-acceleration with an $l=1$ achieves similar performance to a Gaussian beam.
In Fig.~\ref{fig:laguerre} a positron and electron are initialized with $10$ GeV in an $l=1$ mode.
The laser pulse is given the same polarization, duration, beam width $w_0$, and power as the Gaussian pulse shown in Fig.~\ref{fig:gauss}.
As shown in Fig.~\ref{fig:dw_laguerre} the particles achieve a slightly lower $(1+\beta_z)/(\gamma -\gamma\beta_z)$.

The slightly reduced performance, rather than increased re-acceleration, may be understood as follows.
The use of the LG mode is driven by the on axis longitudinal field, but any generated particles will rarely be on the axis.
Particles will be generated in regions where the driving laser and electron beam result in a high $\chi$, which is proportional to the strength of the electromagnetic field in the rest frame of the driving beam.
Longitudinal electric fields don't experience any increase when transformed into this frame, implying particles will be generated within the outer ring of high transverse field.
After particles are generated and decelerated, they may be expelled, or transversely bounce back and forth within the beam.
Only when they move side to side within the beam will they cross the beam axis and be driven by the longitudinal field.
Even when they cross the axis, due to significant transverse momentum they experience the longitudinal field for only a short duration.
Thus, even with the LG mode, the interaction between particles and the longitudinal field is brief, and the transverse field regions provide the dominant source of particle deceleration and re-acceleration.
The dominance of the transverse field in the deceleration and re-acceleration process results in a slightly reduced performance for the LG mode.
This is caused by the lower intensity of  the LG mode as it is spread out over a larger area for equivalent power and beamwidth $w_0$, resulting a lower strength field and force applied to generated particles.

\begin{figure}[htb]
    \includegraphics[trim={0 5 0 0},clip,width=\columnwidth]{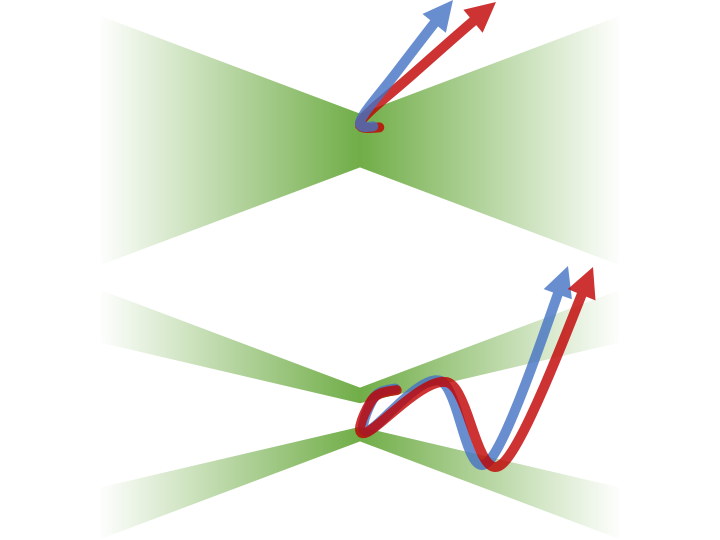}
    \caption{Positrons (red) and electrons (blue) generated in the Gaussian pulse(top) can be quickly scattered. Through hollowing out the beam center an LG mode(bottom) might ponderomotively contain pairs for a longer duration.}
    \label{fig:cartoon_beams}
\end{figure}

The lower strength forcing of the LG mode slightly reduces performance, but a comparison between the LG mode and Gaussian beam demonstrates an advantage for the LG mode in particle confinement.
Both field configurations provide transverse ponderomotive effects, but the LG mode contains regions which can allow each particle to upshift a longer duration of the laser and simultaneously allow more pair density accumulation.
Transverse scattering occurs due to the gradient in electric field strength creating an outward ponderomotive force.
Some outward ponderomotive push will occur for any focussed laser, as the field must trend down in strength away from the focus.
However, in a small region of space the gradient can be reversed through hollowing out the interior of the laser beam and creating a local minimum in field strength.
If particles experience the proper conditions in the high field region, they can be transversely trapped in this ponderomotive well.
In this best of cases this results in longer co-propagation as diagrammed in Fig.~\ref{fig:cartoon_beams}.
Best case behavior results from particles maintaining co-propagation for as long as the well can be maintained, either the pulse duration or the Rayleigh range, instead of the much shorter time it takes to cross a beamwidth of a Gaussian beam.
For typical parameters this can be a significant factor.
Increased particle confinement can be seen in the positron oscillations in Fig.~\ref{fig:dw_laguerre} where the positron experiences one bounce, as shown in Fig.~\ref{fig:laguerre}, and oscillates much longer than the pair in Fig.~\ref{fig:dw_gauss}.
This longer confinement is not guaranteed, and is sensitive to initial conditions, as can be see in the behavior of the electron in Fig.~\ref{fig:laguerre}.

When considering both which type of beam to use and how tightly to focus it, there is a tradeoff between confinement and stopping power.
A tighter focus and Gaussian beam provide more stopping power and re-acceleration, but increase transverse scattering and thus should reduce pair density accumulation.
A wider focus and LG mode provide the opposite balance of these factors.
However, under the right conditions an LG mode may come with minimal cost.
The analytical work in Sec.~\ref{sec:decel_guidelines} suggests that the laser power must primarily exceed a threshold of $a_0 \sim 100$ for pairs to contribute.
If this threshold can be exceeded while the advantages of an LG mode are maintained, additional density accumulation would not tradeoff strongly against per particle contributions.

\section{Summary and Discussion}
\label{sec:sum}
Stopping generated pairs was known to be key for creating large frequency signatures.
In this paper we have demonstrated that not just stopped but re-accelerated particles should drive useful frequency shifts.
Including the effects of re-accelerated particles allows for particles to contribute not only at the point of reflection, but at later times as well.
Re-acceleration changes the criteria by which particles may contribute and, in turn, which laser configurations are favorable for creating detectable signatures.
In a simple plane wave model, particle re-acceleration suggests that each electron or positron could contribute up to a factor of two more than was expected previously.
When quantum synchrotron radiation is considered over multiple laser cycles, we demonstrate that lower laser power can be sufficient to stop electron-positron pairs.
Consequently, if deceleration is the limiting constraint for generating frequency signatures, laser power might be lowered by up to forty percent, simultaneously and advantageously lowering the minimum pair particle energy.

To further magnify frequency changes, we evaluated an LG mode to improve particle re-acceleration, but we found no advantage in particle deceleration and re-acceleration compared against a fundamental mode Gaussian beam.
Unlike in traditional acceleration schemes, the LG beam's longitudinal field doesn't provide additional longitudinal work as the particles are generated, stopped, and driven primarily in the high transverse field ring.
Moreover, when the LG mode is compared to the Gaussian beam, we clarify that not just longitudinal, but transverse forcing is important for generating large signatures.
Transverse forcing becomes important when re-acceleration is considered, as re-accelerated particles confined within the beam can continue to contribute the plasma density.
The structure of the LG mode can transversely confine particles for a longer duration than the Gaussian beam.
Through this comparison, we clarify that Eq.~\eqref{eqn:weighting} is not the sole metric of interest, and to maximize any frequency signature the duration of interaction with the beam should be jointly improved.
Particle contributions after reflection are necessary for longer particle confinement times to be relevant.
That particles may contribute after reflection increases the importance of considering longer durations, and thus the expected enhancement of using an LG mode.

Relaxed requirements of the driving laser combined with the use of an LG mode suggest that signatures can be more easily produced.
We have solidified and expanded previous estimates for the behavior of single particles, but these are only a proxy for estimating the magnitude of the desired collective effects.
Our analysis comes with several caveats.
While the contribution per particle is vital, achieving high density is equally important.
The density of pairs will be determined both by the spatial distribution of their generation, and by whether they experience compression or dispersal during deceleration and reflection, both factors which are primarily neglected here.
The analytical limits we develop in Sec.~\ref{sec:decel_guidelines}, are developed in consideration of pairs with a plane wave.
This should be a fair approximation for the early stages of particle acceleration when the particle primarily moves longitudinally, but will not extend as cleanly when transverse variation is also relevant.
When this is the case, and the electromagnetic potential is no longer only a function of $\phi$, $\gamma(1-\beta_z)$ will no longer be a strictly conserved quantity, and this is a noticeable effect for $\omega t > 0$ in Fig.~\ref{fig:dw}.
Additional complications from changes of the driving beam by pairs, as well as pair-pair interactions are also worthy of consideration and may change particle dynamics.

To address uncertainties and further maximize signature strength, there are multiple immediate directions for future work.
Particle in cell simulations at lower beam powers with LG mode driving beams could validate predictions of improved performance.
When considering the full range of collective effects, not only using a LG mode, but a radially polarized LG mode, might further improve performance by forcing radial symmetry and allowing compression in the radial direction of generated particles.
The radiation reaction in the later stages of particle acceleration can also substantially change the dynamics of particle acceleration.\cite{tamburini_radiation_2010, chen_radiation_2010, mishra_effect_2021, mishra_exact_2021, yeh_strong_2021}
Further optimization of its role in the later stages of the particle dynamics could improve particle confinement and re-acceleration.
If the effective plasma frequency is further magnified, spectral information could serve as an even more effective signature of collective QED effects.

\begin{acknowledgments}
    This work was supported by Grants DENA0003871 and NNSA DE-SC0021248.
\end{acknowledgments}

\bibliography{decel_bib}

\begin{thebibliography}{40}%
\makeatletter
\providecommand \@ifxundefined [1]{%
 \@ifx{#1\undefined}
}%
\providecommand \@ifnum [1]{%
 \ifnum #1\expandafter \@firstoftwo
 \else \expandafter \@secondoftwo
 \fi
}%
\providecommand \@ifx [1]{%
 \ifx #1\expandafter \@firstoftwo
 \else \expandafter \@secondoftwo
 \fi
}%
\providecommand \natexlab [1]{#1}%
\providecommand \enquote  [1]{``#1''}%
\providecommand \bibnamefont  [1]{#1}%
\providecommand \bibfnamefont [1]{#1}%
\providecommand \citenamefont [1]{#1}%
\providecommand \href@noop [0]{\@secondoftwo}%
\providecommand \href [0]{\begingroup \@sanitize@url \@href}%
\providecommand \@href[1]{\@@startlink{#1}\@@href}%
\providecommand \@@href[1]{\endgroup#1\@@endlink}%
\providecommand \@sanitize@url [0]{\catcode `\\12\catcode `\$12\catcode
  `\&12\catcode `\#12\catcode `\^12\catcode `\_12\catcode `\%12\relax}%
\providecommand \@@startlink[1]{}%
\providecommand \@@endlink[0]{}%
\providecommand \url  [0]{\begingroup\@sanitize@url \@url }%
\providecommand \@url [1]{\endgroup\@href {#1}{\urlprefix }}%
\providecommand \urlprefix  [0]{URL }%
\providecommand \Eprint [0]{\href }%
\providecommand \doibase [0]{http://dx.doi.org/}%
\providecommand \selectlanguage [0]{\@gobble}%
\providecommand \bibinfo  [0]{\@secondoftwo}%
\providecommand \bibfield  [0]{\@secondoftwo}%
\providecommand \translation [1]{[#1]}%
\providecommand \BibitemOpen [0]{}%
\providecommand \bibitemStop [0]{}%
\providecommand \bibitemNoStop [0]{.\EOS\space}%
\providecommand \EOS [0]{\spacefactor3000\relax}%
\providecommand \BibitemShut  [1]{\csname bibitem#1\endcsname}%
\let\auto@bib@innerbib\@empty
\bibitem [{\citenamefont {Bell}\ and\ \citenamefont
  {Kirk}(2008)}]{bell_possibility_2008}%
  \BibitemOpen
  \bibfield  {author} {\bibinfo {author} {\bibfnamefont {A.~R.}\ \bibnamefont
  {Bell}}\ and\ \bibinfo {author} {\bibfnamefont {J.~G.}\ \bibnamefont
  {Kirk}},\ }\bibfield  {title} {\enquote {\bibinfo {title} {Possibility of
  {{Prolific Pair Production}} with {{High-Power Lasers}}},}\ }\href {\doibase
  10.1103/PhysRevLett.101.200403} {\bibfield  {journal} {\bibinfo  {journal}
  {Physical Review Letters}\ }\textbf {\bibinfo {volume} {101}},\ \bibinfo
  {pages} {200403} (\bibinfo {year} {2008})}\BibitemShut {NoStop}%
\bibitem [{\citenamefont {Fedotov}\ \emph {et~al.}(2010)\citenamefont
  {Fedotov}, \citenamefont {Narozhny}, \citenamefont {Mourou},\ and\
  \citenamefont {Korn}}]{fedotov_limitations_2010}%
  \BibitemOpen
  \bibfield  {author} {\bibinfo {author} {\bibfnamefont {A.~M.}\ \bibnamefont
  {Fedotov}}, \bibinfo {author} {\bibfnamefont {N.~B.}\ \bibnamefont
  {Narozhny}}, \bibinfo {author} {\bibfnamefont {G.}~\bibnamefont {Mourou}}, \
  and\ \bibinfo {author} {\bibfnamefont {G.}~\bibnamefont {Korn}},\ }\bibfield
  {title} {\enquote {\bibinfo {title} {Limitations on the {{Attainable
  Intensity}} of {{High Power Lasers}}},}\ }\href {\doibase
  10.1103/PhysRevLett.105.080402} {\bibfield  {journal} {\bibinfo  {journal}
  {Physical Review Letters}\ }\textbf {\bibinfo {volume} {105}},\ \bibinfo
  {pages} {080402} (\bibinfo {year} {2010})}\BibitemShut {NoStop}%
\bibitem [{\citenamefont {Bulanov}\ \emph {et~al.}(2010)\citenamefont
  {Bulanov}, \citenamefont {Esirkepov}, \citenamefont {Thomas}, \citenamefont
  {Koga},\ and\ \citenamefont {Bulanov}}]{bulanov_schwinger_2010}%
  \BibitemOpen
  \bibfield  {author} {\bibinfo {author} {\bibfnamefont {S.~S.}\ \bibnamefont
  {Bulanov}}, \bibinfo {author} {\bibfnamefont {T.~Z.}\ \bibnamefont
  {Esirkepov}}, \bibinfo {author} {\bibfnamefont {A.~G.~R.}\ \bibnamefont
  {Thomas}}, \bibinfo {author} {\bibfnamefont {J.~K.}\ \bibnamefont {Koga}}, \
  and\ \bibinfo {author} {\bibfnamefont {S.~V.}\ \bibnamefont {Bulanov}},\
  }\bibfield  {title} {\enquote {\bibinfo {title} {Schwinger {{Limit
  Attainability}} with {{Extreme Power Lasers}}},}\ }\href {\doibase
  10.1103/PhysRevLett.105.220407} {\bibfield  {journal} {\bibinfo  {journal}
  {Physical Review Letters}\ }\textbf {\bibinfo {volume} {105}},\ \bibinfo
  {pages} {220407} (\bibinfo {year} {2010})}\BibitemShut {NoStop}%
\bibitem [{\citenamefont {Zhu}\ \emph {et~al.}(2016)\citenamefont {Zhu},
  \citenamefont {Yu}, \citenamefont {Sheng}, \citenamefont {Yin}, \citenamefont
  {Turcu},\ and\ \citenamefont {Pukhov}}]{zhu_dense_2016}%
  \BibitemOpen
  \bibfield  {author} {\bibinfo {author} {\bibfnamefont {X.-L.}\ \bibnamefont
  {Zhu}}, \bibinfo {author} {\bibfnamefont {T.-P.}\ \bibnamefont {Yu}},
  \bibinfo {author} {\bibfnamefont {Z.-M.}\ \bibnamefont {Sheng}}, \bibinfo
  {author} {\bibfnamefont {Y.}~\bibnamefont {Yin}}, \bibinfo {author}
  {\bibfnamefont {I.~C.~E.}\ \bibnamefont {Turcu}}, \ and\ \bibinfo {author}
  {\bibfnamefont {A.}~\bibnamefont {Pukhov}},\ }\bibfield  {title} {\enquote
  {\bibinfo {title} {Dense {{GeV}} electron\textendash positron pairs generated
  by lasers in near-critical-density plasmas},}\ }\href {\doibase
  10.1038/ncomms13686} {\bibfield  {journal} {\bibinfo  {journal} {Nature
  Communications}\ }\textbf {\bibinfo {volume} {7}},\ \bibinfo {pages} {13686}
  (\bibinfo {year} {2016})}\BibitemShut {NoStop}%
\bibitem [{\citenamefont {Yakimenko}\ \emph {et~al.}(2019)\citenamefont
  {Yakimenko}, \citenamefont {Meuren}, \citenamefont {Del~Gaudio},
  \citenamefont {Baumann}, \citenamefont {Fedotov}, \citenamefont {Fiuza},
  \citenamefont {Grismayer}, \citenamefont {Hogan}, \citenamefont {Pukhov},
  \citenamefont {Silva},\ and\ \citenamefont
  {White}}]{yakimenko_prospect_2019}%
  \BibitemOpen
  \bibfield  {author} {\bibinfo {author} {\bibfnamefont {V.}~\bibnamefont
  {Yakimenko}}, \bibinfo {author} {\bibfnamefont {S.}~\bibnamefont {Meuren}},
  \bibinfo {author} {\bibfnamefont {F.}~\bibnamefont {Del~Gaudio}}, \bibinfo
  {author} {\bibfnamefont {C.}~\bibnamefont {Baumann}}, \bibinfo {author}
  {\bibfnamefont {A.}~\bibnamefont {Fedotov}}, \bibinfo {author} {\bibfnamefont
  {F.}~\bibnamefont {Fiuza}}, \bibinfo {author} {\bibfnamefont
  {T.}~\bibnamefont {Grismayer}}, \bibinfo {author} {\bibfnamefont {M.~J.}\
  \bibnamefont {Hogan}}, \bibinfo {author} {\bibfnamefont {A.}~\bibnamefont
  {Pukhov}}, \bibinfo {author} {\bibfnamefont {L.~O.}\ \bibnamefont {Silva}}, \
  and\ \bibinfo {author} {\bibfnamefont {G.}~\bibnamefont {White}},\ }\bibfield
   {title} {\enquote {\bibinfo {title} {Prospect of {{Studying Nonperturbative
  QED}} with {{Beam-Beam Collisions}}},}\ }\href {\doibase
  10.1103/PhysRevLett.122.190404} {\bibfield  {journal} {\bibinfo  {journal}
  {Physical Review Letters}\ }\textbf {\bibinfo {volume} {122}},\ \bibinfo
  {pages} {190404} (\bibinfo {year} {2019})}\BibitemShut {NoStop}%
\bibitem [{\citenamefont {Di~Piazza}\ \emph {et~al.}(2012)\citenamefont
  {Di~Piazza}, \citenamefont {M{\"u}ller}, \citenamefont {Hatsagortsyan},\ and\
  \citenamefont {Keitel}}]{di_piazza_extremely_2012}%
  \BibitemOpen
  \bibfield  {author} {\bibinfo {author} {\bibfnamefont {A.}~\bibnamefont
  {Di~Piazza}}, \bibinfo {author} {\bibfnamefont {C.}~\bibnamefont
  {M{\"u}ller}}, \bibinfo {author} {\bibfnamefont {K.~Z.}\ \bibnamefont
  {Hatsagortsyan}}, \ and\ \bibinfo {author} {\bibfnamefont {C.~H.}\
  \bibnamefont {Keitel}},\ }\bibfield  {title} {\enquote {\bibinfo {title}
  {Extremely high-intensity laser interactions with fundamental quantum
  systems},}\ }\href {\doibase 10.1103/RevModPhys.84.1177} {\bibfield
  {journal} {\bibinfo  {journal} {Reviews of Modern Physics}\ }\textbf
  {\bibinfo {volume} {84}},\ \bibinfo {pages} {1177--1228} (\bibinfo {year}
  {2012})}\BibitemShut {NoStop}%
\bibitem [{\citenamefont {Nerush}\ \emph {et~al.}(2011)\citenamefont {Nerush},
  \citenamefont {Kostyukov}, \citenamefont {Fedotov}, \citenamefont {Narozhny},
  \citenamefont {Elkina},\ and\ \citenamefont {Ruhl}}]{nerush_laser_2011}%
  \BibitemOpen
  \bibfield  {author} {\bibinfo {author} {\bibfnamefont {E.~N.}\ \bibnamefont
  {Nerush}}, \bibinfo {author} {\bibfnamefont {I.~Y.}\ \bibnamefont
  {Kostyukov}}, \bibinfo {author} {\bibfnamefont {A.~M.}\ \bibnamefont
  {Fedotov}}, \bibinfo {author} {\bibfnamefont {N.~B.}\ \bibnamefont
  {Narozhny}}, \bibinfo {author} {\bibfnamefont {N.~V.}\ \bibnamefont
  {Elkina}}, \ and\ \bibinfo {author} {\bibfnamefont {H.}~\bibnamefont
  {Ruhl}},\ }\bibfield  {title} {\enquote {\bibinfo {title} {Laser {{Field
  Absorption}} in {{Self-Generated Electron-Positron Pair Plasma}}},}\ }\href
  {\doibase 10.1103/PhysRevLett.106.035001} {\bibfield  {journal} {\bibinfo
  {journal} {Physical Review Letters}\ }\textbf {\bibinfo {volume} {106}},\
  \bibinfo {pages} {035001} (\bibinfo {year} {2011})}\BibitemShut {NoStop}%
\bibitem [{\citenamefont {Grismayer}\ \emph {et~al.}(2016)\citenamefont
  {Grismayer}, \citenamefont {Vranic}, \citenamefont {Martins}, \citenamefont
  {Fonseca},\ and\ \citenamefont {Silva}}]{grismayer_laser_2016}%
  \BibitemOpen
  \bibfield  {author} {\bibinfo {author} {\bibfnamefont {T.}~\bibnamefont
  {Grismayer}}, \bibinfo {author} {\bibfnamefont {M.}~\bibnamefont {Vranic}},
  \bibinfo {author} {\bibfnamefont {J.~L.}\ \bibnamefont {Martins}}, \bibinfo
  {author} {\bibfnamefont {R.~A.}\ \bibnamefont {Fonseca}}, \ and\ \bibinfo
  {author} {\bibfnamefont {L.~O.}\ \bibnamefont {Silva}},\ }\bibfield  {title}
  {\enquote {\bibinfo {title} {Laser absorption via quantum electrodynamics
  cascades in counter propagating laser pulses},}\ }\href {\doibase
  10.1063/1.4950841} {\bibfield  {journal} {\bibinfo  {journal} {Physics of
  Plasmas}\ }\textbf {\bibinfo {volume} {23}},\ \bibinfo {pages} {056706}
  (\bibinfo {year} {2016})}\BibitemShut {NoStop}%
\bibitem [{\citenamefont {Savin}\ \emph {et~al.}(2019)\citenamefont {Savin},
  \citenamefont {Ross}, \citenamefont {Aboushelbaya}, \citenamefont {Mayr},
  \citenamefont {Spiers}, \citenamefont {Wang},\ and\ \citenamefont
  {Norreys}}]{savin_energy_2019}%
  \BibitemOpen
  \bibfield  {author} {\bibinfo {author} {\bibfnamefont {A.~F.}\ \bibnamefont
  {Savin}}, \bibinfo {author} {\bibfnamefont {A.~J.}\ \bibnamefont {Ross}},
  \bibinfo {author} {\bibfnamefont {R.}~\bibnamefont {Aboushelbaya}}, \bibinfo
  {author} {\bibfnamefont {M.~W.}\ \bibnamefont {Mayr}}, \bibinfo {author}
  {\bibfnamefont {B.}~\bibnamefont {Spiers}}, \bibinfo {author} {\bibfnamefont
  {R.~H.-W.}\ \bibnamefont {Wang}}, \ and\ \bibinfo {author} {\bibfnamefont
  {P.~A.}\ \bibnamefont {Norreys}},\ }\bibfield  {title} {\enquote {\bibinfo
  {title} {Energy absorption in the laser-{{QED}} regime},}\ }\href {\doibase
  10.1038/s41598-019-45536-x} {\bibfield  {journal} {\bibinfo  {journal}
  {Scientific Reports}\ }\textbf {\bibinfo {volume} {9}},\ \bibinfo {pages}
  {8956} (\bibinfo {year} {2019})}\BibitemShut {NoStop}%
\bibitem [{\citenamefont {Edwards}, \citenamefont {Fisch},\ and\ \citenamefont
  {Mikhailova}(2016)}]{edwards_strongly_2016}%
  \BibitemOpen
  \bibfield  {author} {\bibinfo {author} {\bibfnamefont {M.~R.}\ \bibnamefont
  {Edwards}}, \bibinfo {author} {\bibfnamefont {N.~J.}\ \bibnamefont {Fisch}},
  \ and\ \bibinfo {author} {\bibfnamefont {J.~M.}\ \bibnamefont {Mikhailova}},\
  }\bibfield  {title} {\enquote {\bibinfo {title} {Strongly {{Enhanced
  Stimulated Brillouin Backscattering}} in an {{Electron-Positron Plasma}}},}\
  }\href {\doibase 10.1103/PhysRevLett.116.015004} {\bibfield  {journal}
  {\bibinfo  {journal} {Physical Review Letters}\ }\textbf {\bibinfo {volume}
  {116}},\ \bibinfo {pages} {015004} (\bibinfo {year} {2016})}\BibitemShut
  {NoStop}%
\bibitem [{\citenamefont {Schluck}, \citenamefont {Lehmann},\ and\
  \citenamefont {Spatschek}(2017)}]{schluck_parametric_2017}%
  \BibitemOpen
  \bibfield  {author} {\bibinfo {author} {\bibfnamefont {F.}~\bibnamefont
  {Schluck}}, \bibinfo {author} {\bibfnamefont {G.}~\bibnamefont {Lehmann}}, \
  and\ \bibinfo {author} {\bibfnamefont {K.~H.}\ \bibnamefont {Spatschek}},\
  }\bibfield  {title} {\enquote {\bibinfo {title} {Parametric pulse
  amplification by acoustic quasimodes in electron-positron plasma},}\ }\href
  {\doibase 10.1103/PhysRevE.96.053204} {\bibfield  {journal} {\bibinfo
  {journal} {Physical Review E}\ }\textbf {\bibinfo {volume} {96}},\ \bibinfo
  {pages} {053204} (\bibinfo {year} {2017})}\BibitemShut {NoStop}%
\bibitem [{\citenamefont {Gong}\ \emph {et~al.}(2018)\citenamefont {Gong},
  \citenamefont {Hu}, \citenamefont {Lu}, \citenamefont {Yu}, \citenamefont
  {Wang}, \citenamefont {Fu}, \citenamefont {Chen}, \citenamefont {He},\ and\
  \citenamefont {Yan}}]{gong_brilliant_2018}%
  \BibitemOpen
  \bibfield  {author} {\bibinfo {author} {\bibfnamefont {Z.}~\bibnamefont
  {Gong}}, \bibinfo {author} {\bibfnamefont {R.~H.}\ \bibnamefont {Hu}},
  \bibinfo {author} {\bibfnamefont {H.~Y.}\ \bibnamefont {Lu}}, \bibinfo
  {author} {\bibfnamefont {J.~Q.}\ \bibnamefont {Yu}}, \bibinfo {author}
  {\bibfnamefont {D.~H.}\ \bibnamefont {Wang}}, \bibinfo {author}
  {\bibfnamefont {E.~G.}\ \bibnamefont {Fu}}, \bibinfo {author} {\bibfnamefont
  {C.~E.}\ \bibnamefont {Chen}}, \bibinfo {author} {\bibfnamefont {X.~T.}\
  \bibnamefont {He}}, \ and\ \bibinfo {author} {\bibfnamefont {X.~Q.}\
  \bibnamefont {Yan}},\ }\bibfield  {title} {\enquote {\bibinfo {title}
  {Brilliant {{GeV}} gamma-ray flash from inverse {{Compton}} scattering in the
  {{QED}} regime},}\ }\href {\doibase 10.1088/1361-6587/aaa9b1} {\bibfield
  {journal} {\bibinfo  {journal} {Plasma Physics and Controlled Fusion}\
  }\textbf {\bibinfo {volume} {60}},\ \bibinfo {pages} {044004} (\bibinfo
  {year} {2018})}\BibitemShut {NoStop}%
\bibitem [{\citenamefont {Tiwary}\ and\ \citenamefont
  {Kumar}(2021)}]{tiwary_particle_2021}%
  \BibitemOpen
  \bibfield  {author} {\bibinfo {author} {\bibfnamefont {S.}~\bibnamefont
  {Tiwary}}\ and\ \bibinfo {author} {\bibfnamefont {N.}~\bibnamefont {Kumar}},\
  }\bibfield  {title} {\enquote {\bibinfo {title} {Particle jets in colliding
  two ultraintense laser pulses of varying frequencies},}\ }\href {\doibase
  10.1103/PhysRevResearch.3.043190} {\bibfield  {journal} {\bibinfo  {journal}
  {Physical Review Research}\ }\textbf {\bibinfo {volume} {3}},\ \bibinfo
  {pages} {043190} (\bibinfo {year} {2021})}\BibitemShut {NoStop}%
\bibitem [{\citenamefont {Huang}\ \emph {et~al.}(2021)\citenamefont {Huang},
  \citenamefont {Weng}, \citenamefont {Zhu}, \citenamefont {Li}, \citenamefont
  {Chen}, \citenamefont {Murakami},\ and\ \citenamefont
  {Sheng}}]{huang_relativistic-induced_2021}%
  \BibitemOpen
  \bibfield  {author} {\bibinfo {author} {\bibfnamefont {J.}~\bibnamefont
  {Huang}}, \bibinfo {author} {\bibfnamefont {S.~M.}\ \bibnamefont {Weng}},
  \bibinfo {author} {\bibfnamefont {X.~L.}\ \bibnamefont {Zhu}}, \bibinfo
  {author} {\bibfnamefont {X.~F.}\ \bibnamefont {Li}}, \bibinfo {author}
  {\bibfnamefont {M.}~\bibnamefont {Chen}}, \bibinfo {author} {\bibfnamefont
  {M.}~\bibnamefont {Murakami}}, \ and\ \bibinfo {author} {\bibfnamefont
  {Z.~M.}\ \bibnamefont {Sheng}},\ }\bibfield  {title} {\enquote {\bibinfo
  {title} {Relativistic-induced opacity of electron\textendash positron
  plasmas},}\ }\href {\doibase 10.1088/1361-6587/abe0f9} {\bibfield  {journal}
  {\bibinfo  {journal} {Plasma Physics and Controlled Fusion}\ }\textbf
  {\bibinfo {volume} {63}},\ \bibinfo {pages} {045010} (\bibinfo {year}
  {2021})}\BibitemShut {NoStop}%
\bibitem [{\citenamefont {Li}\ \emph {et~al.}(2020)\citenamefont {Li},
  \citenamefont {Chen}, \citenamefont {Wang},\ and\ \citenamefont
  {Hu}}]{li_production_2020}%
  \BibitemOpen
  \bibfield  {author} {\bibinfo {author} {\bibfnamefont {Y.-F.}\ \bibnamefont
  {Li}}, \bibinfo {author} {\bibfnamefont {Y.-Y.}\ \bibnamefont {Chen}},
  \bibinfo {author} {\bibfnamefont {W.-M.}\ \bibnamefont {Wang}}, \ and\
  \bibinfo {author} {\bibfnamefont {H.-S.}\ \bibnamefont {Hu}},\ }\bibfield
  {title} {\enquote {\bibinfo {title} {Production of {{Highly Polarized
  Positron Beams}} via {{Helicity Transfer}} from {{Polarized Electrons}} in a
  {{Strong Laser Field}}},}\ }\href {\doibase 10.1103/PhysRevLett.125.044802}
  {\bibfield  {journal} {\bibinfo  {journal} {Physical Review Letters}\
  }\textbf {\bibinfo {volume} {125}},\ \bibinfo {pages} {044802} (\bibinfo
  {year} {2020})}\BibitemShut {NoStop}%
\bibitem [{\citenamefont {Qu}, \citenamefont {Meuren},\ and\ \citenamefont
  {Fisch}(2021{\natexlab{a}})}]{qu_signature_2021}%
  \BibitemOpen
  \bibfield  {author} {\bibinfo {author} {\bibfnamefont {K.}~\bibnamefont
  {Qu}}, \bibinfo {author} {\bibfnamefont {S.}~\bibnamefont {Meuren}}, \ and\
  \bibinfo {author} {\bibfnamefont {N.~J.}\ \bibnamefont {Fisch}},\ }\bibfield
  {title} {\enquote {\bibinfo {title} {Signature of {{Collective Plasma
  Effects}} in {{Beam-Driven QED Cascades}}},}\ }\href {\doibase
  10.1103/PhysRevLett.127.095001} {\bibfield  {journal} {\bibinfo  {journal}
  {Physical Review Letters}\ }\textbf {\bibinfo {volume} {127}},\ \bibinfo
  {pages} {095001} (\bibinfo {year} {2021}{\natexlab{a}})}\BibitemShut
  {NoStop}%
\bibitem [{\citenamefont {Qu}, \citenamefont {Meuren},\ and\ \citenamefont
  {Fisch}(2021{\natexlab{b}})}]{qu_collective_2021}%
  \BibitemOpen
  \bibfield  {author} {\bibinfo {author} {\bibfnamefont {K.}~\bibnamefont
  {Qu}}, \bibinfo {author} {\bibfnamefont {S.}~\bibnamefont {Meuren}}, \ and\
  \bibinfo {author} {\bibfnamefont {N.~J.}\ \bibnamefont {Fisch}},\ }\bibfield
  {title} {\enquote {\bibinfo {title} {Collective plasma effects of
  electron-positron pairs in beam-driven {{QED}} cascades},}\ }\href@noop {}
  {\bibfield  {journal} {\bibinfo  {journal} {arXiv:2110.12592 [physics]}\ }
  (\bibinfo {year} {2021}{\natexlab{b}})},\ \Eprint
  {http://arxiv.org/abs/2110.12592} {arXiv:2110.12592 [physics]} \BibitemShut
  {NoStop}%
\bibitem [{\citenamefont {Meuren}\ \emph {et~al.}(2020)\citenamefont {Meuren},
  \citenamefont {Bucksbaum}, \citenamefont {Fisch}, \citenamefont {Fi{\'u}za},
  \citenamefont {Glenzer}, \citenamefont {Hogan}, \citenamefont {Qu},
  \citenamefont {Reis}, \citenamefont {White},\ and\ \citenamefont
  {Yakimenko}}]{meuren_seminal_2020-1}%
  \BibitemOpen
  \bibfield  {author} {\bibinfo {author} {\bibfnamefont {S.}~\bibnamefont
  {Meuren}}, \bibinfo {author} {\bibfnamefont {P.~H.}\ \bibnamefont
  {Bucksbaum}}, \bibinfo {author} {\bibfnamefont {N.~J.}\ \bibnamefont
  {Fisch}}, \bibinfo {author} {\bibfnamefont {F.}~\bibnamefont {Fi{\'u}za}},
  \bibinfo {author} {\bibfnamefont {S.}~\bibnamefont {Glenzer}}, \bibinfo
  {author} {\bibfnamefont {M.~J.}\ \bibnamefont {Hogan}}, \bibinfo {author}
  {\bibfnamefont {K.}~\bibnamefont {Qu}}, \bibinfo {author} {\bibfnamefont
  {D.~A.}\ \bibnamefont {Reis}}, \bibinfo {author} {\bibfnamefont
  {G.}~\bibnamefont {White}}, \ and\ \bibinfo {author} {\bibfnamefont
  {V.}~\bibnamefont {Yakimenko}},\ }\bibfield  {title} {\enquote {\bibinfo
  {title} {On {{Seminal HEDP Research Opportunities Enabled}} by {{Colocating
  Multi-Petawatt Laser}} with {{High-Density Electron Beams}}},}\ }\href@noop
  {} {\bibfield  {journal} {\bibinfo  {journal} {arXiv:2002.10051 [hep-ph,
  physics:physics]}\ } (\bibinfo {year} {2020})},\ \Eprint
  {http://arxiv.org/abs/2002.10051} {arXiv:2002.10051 [hep-ph,
  physics:physics]} \BibitemShut {NoStop}%
\bibitem [{\citenamefont {Meuren}\ \emph {et~al.}(2021)\citenamefont {Meuren},
  \citenamefont {Reis}, \citenamefont {Blandford}, \citenamefont {Bucksbaum},
  \citenamefont {Fisch}, \citenamefont {Fiuza}, \citenamefont {Gerstmayr},
  \citenamefont {Glenzer}, \citenamefont {Hogan}, \citenamefont {Pellegrini},
  \citenamefont {Peskin}, \citenamefont {Qu}, \citenamefont {White},\ and\
  \citenamefont {Yakimenko}}]{meuren_mp3_2021}%
  \BibitemOpen
  \bibfield  {author} {\bibinfo {author} {\bibfnamefont {S.}~\bibnamefont
  {Meuren}}, \bibinfo {author} {\bibfnamefont {D.~A.}\ \bibnamefont {Reis}},
  \bibinfo {author} {\bibfnamefont {R.}~\bibnamefont {Blandford}}, \bibinfo
  {author} {\bibfnamefont {P.~H.}\ \bibnamefont {Bucksbaum}}, \bibinfo {author}
  {\bibfnamefont {N.~J.}\ \bibnamefont {Fisch}}, \bibinfo {author}
  {\bibfnamefont {F.}~\bibnamefont {Fiuza}}, \bibinfo {author} {\bibfnamefont
  {E.}~\bibnamefont {Gerstmayr}}, \bibinfo {author} {\bibfnamefont
  {S.}~\bibnamefont {Glenzer}}, \bibinfo {author} {\bibfnamefont {M.~J.}\
  \bibnamefont {Hogan}}, \bibinfo {author} {\bibfnamefont {C.}~\bibnamefont
  {Pellegrini}}, \bibinfo {author} {\bibfnamefont {M.~E.}\ \bibnamefont
  {Peskin}}, \bibinfo {author} {\bibfnamefont {K.}~\bibnamefont {Qu}}, \bibinfo
  {author} {\bibfnamefont {G.}~\bibnamefont {White}}, \ and\ \bibinfo {author}
  {\bibfnamefont {V.}~\bibnamefont {Yakimenko}},\ }\bibfield  {title} {\enquote
  {\bibinfo {title} {{{MP3 White Paper}} 2021 -- {{Research Opportunities
  Enabled}} by {{Co-locating Multi-Petawatt Lasers}} with {{Dense
  Ultra-Relativistic Electron Beams}}},}\ }\href@noop {} {\bibfield  {journal}
  {\bibinfo  {journal} {arXiv:2105.11607 [hep-ph, physics:physics]}\ }
  (\bibinfo {year} {2021})},\ \Eprint {http://arxiv.org/abs/2105.11607}
  {arXiv:2105.11607 [hep-ph, physics:physics]} \BibitemShut {NoStop}%
\bibitem [{\citenamefont {Amaro}\ and\ \citenamefont
  {Vranic}(2021)}]{amaro_optimal_2021}%
  \BibitemOpen
  \bibfield  {author} {\bibinfo {author} {\bibfnamefont {{\'O}.}~\bibnamefont
  {Amaro}}\ and\ \bibinfo {author} {\bibfnamefont {M.}~\bibnamefont {Vranic}},\
  }\bibfield  {title} {\enquote {\bibinfo {title} {Optimal laser focusing for
  positron production in laser\textendash electron scattering},}\ }\href
  {\doibase 10.1088/1367-2630/ac2e83} {\bibfield  {journal} {\bibinfo
  {journal} {New Journal of Physics}\ }\textbf {\bibinfo {volume} {23}},\
  \bibinfo {pages} {115001} (\bibinfo {year} {2021})}\BibitemShut {NoStop}%
\bibitem [{\citenamefont {Wood}, \citenamefont {Siders},\ and\ \citenamefont
  {Downer}(1991)}]{wood_measurement_1991}%
  \BibitemOpen
  \bibfield  {author} {\bibinfo {author} {\bibfnamefont {W.~M.}\ \bibnamefont
  {Wood}}, \bibinfo {author} {\bibfnamefont {C.~W.}\ \bibnamefont {Siders}}, \
  and\ \bibinfo {author} {\bibfnamefont {M.~C.}\ \bibnamefont {Downer}},\
  }\bibfield  {title} {\enquote {\bibinfo {title} {Measurement of femtosecond
  ionization dynamics of atmospheric density gases by spectral blueshifting},}\
  }\href {\doibase 10.1103/PhysRevLett.67.3523} {\bibfield  {journal} {\bibinfo
   {journal} {Physical Review Letters}\ }\textbf {\bibinfo {volume} {67}},\
  \bibinfo {pages} {3523--3526} (\bibinfo {year} {1991})}\BibitemShut {NoStop}%
\bibitem [{\citenamefont {Nishida}\ \emph {et~al.}(2012)\citenamefont
  {Nishida}, \citenamefont {Yugami}, \citenamefont {Higashiguchi},
  \citenamefont {Otsuka}, \citenamefont {Suzuki}, \citenamefont {Nakata},
  \citenamefont {Sentoku},\ and\ \citenamefont
  {Kodama}}]{nishida_experimental_2012}%
  \BibitemOpen
  \bibfield  {author} {\bibinfo {author} {\bibfnamefont {A.}~\bibnamefont
  {Nishida}}, \bibinfo {author} {\bibfnamefont {N.}~\bibnamefont {Yugami}},
  \bibinfo {author} {\bibfnamefont {T.}~\bibnamefont {Higashiguchi}}, \bibinfo
  {author} {\bibfnamefont {T.}~\bibnamefont {Otsuka}}, \bibinfo {author}
  {\bibfnamefont {F.}~\bibnamefont {Suzuki}}, \bibinfo {author} {\bibfnamefont
  {M.}~\bibnamefont {Nakata}}, \bibinfo {author} {\bibfnamefont
  {Y.}~\bibnamefont {Sentoku}}, \ and\ \bibinfo {author} {\bibfnamefont
  {R.}~\bibnamefont {Kodama}},\ }\bibfield  {title} {\enquote {\bibinfo {title}
  {Experimental observation of frequency up-conversion by flash ionization},}\
  }\href {\doibase 10.1063/1.4755843} {\bibfield  {journal} {\bibinfo
  {journal} {Applied Physics Letters}\ }\textbf {\bibinfo {volume} {101}},\
  \bibinfo {pages} {161118} (\bibinfo {year} {2012})}\BibitemShut {NoStop}%
\bibitem [{\citenamefont {Qu}\ and\ \citenamefont
  {Fisch}(2019)}]{qu_laser_2019}%
  \BibitemOpen
  \bibfield  {author} {\bibinfo {author} {\bibfnamefont {K.}~\bibnamefont
  {Qu}}\ and\ \bibinfo {author} {\bibfnamefont {N.~J.}\ \bibnamefont {Fisch}},\
  }\bibfield  {title} {\enquote {\bibinfo {title} {Laser frequency upconversion
  in plasmas with finite ionization rates},}\ }\href {\doibase
  10.1063/1.5110292} {\bibfield  {journal} {\bibinfo  {journal} {Physics of
  Plasmas}\ }\textbf {\bibinfo {volume} {26}},\ \bibinfo {pages} {083105}
  (\bibinfo {year} {2019})}\BibitemShut {NoStop}%
\bibitem [{\citenamefont {Hartemann}\ \emph {et~al.}(1995)\citenamefont
  {Hartemann}, \citenamefont {Fochs}, \citenamefont {Le~Sage}, \citenamefont
  {Luhmann}, \citenamefont {Woodworth}, \citenamefont {Perry}, \citenamefont
  {Chen},\ and\ \citenamefont {Kerman}}]{hartemann_nonlinear_1995}%
  \BibitemOpen
  \bibfield  {author} {\bibinfo {author} {\bibfnamefont {F.~V.}\ \bibnamefont
  {Hartemann}}, \bibinfo {author} {\bibfnamefont {S.~N.}\ \bibnamefont
  {Fochs}}, \bibinfo {author} {\bibfnamefont {G.~P.}\ \bibnamefont {Le~Sage}},
  \bibinfo {author} {\bibfnamefont {N.~C.}\ \bibnamefont {Luhmann}}, \bibinfo
  {author} {\bibfnamefont {J.~G.}\ \bibnamefont {Woodworth}}, \bibinfo {author}
  {\bibfnamefont {M.~D.}\ \bibnamefont {Perry}}, \bibinfo {author}
  {\bibfnamefont {Y.~J.}\ \bibnamefont {Chen}}, \ and\ \bibinfo {author}
  {\bibfnamefont {A.~K.}\ \bibnamefont {Kerman}},\ }\bibfield  {title}
  {\enquote {\bibinfo {title} {Nonlinear ponderomotive scattering of
  relativistic electrons by an intense laser field at focus},}\ }\href
  {\doibase 10.1103/PhysRevE.51.4833} {\bibfield  {journal} {\bibinfo
  {journal} {Physical Review E}\ }\textbf {\bibinfo {volume} {51}},\ \bibinfo
  {pages} {4833--4843} (\bibinfo {year} {1995})}\BibitemShut {NoStop}%
\bibitem [{\citenamefont {Blackburn}(2020)}]{blackburn_radiation_2020}%
  \BibitemOpen
  \bibfield  {author} {\bibinfo {author} {\bibfnamefont {T.~G.}\ \bibnamefont
  {Blackburn}},\ }\bibfield  {title} {\enquote {\bibinfo {title} {Radiation
  reaction in electron\textendash beam interactions with high-intensity
  lasers},}\ }\href {\doibase 10.1007/s41614-020-0042-0} {\bibfield  {journal}
  {\bibinfo  {journal} {Reviews of Modern Plasma Physics}\ }\textbf {\bibinfo
  {volume} {4}},\ \bibinfo {pages} {5} (\bibinfo {year} {2020})}\BibitemShut
  {NoStop}%
\bibitem [{\citenamefont {Tamburini}, \citenamefont {Di~Piazza},\ and\
  \citenamefont {Keitel}(2017)}]{tamburini_laser-pulse-shape_2017}%
  \BibitemOpen
  \bibfield  {author} {\bibinfo {author} {\bibfnamefont {M.}~\bibnamefont
  {Tamburini}}, \bibinfo {author} {\bibfnamefont {A.}~\bibnamefont
  {Di~Piazza}}, \ and\ \bibinfo {author} {\bibfnamefont {C.~H.}\ \bibnamefont
  {Keitel}},\ }\bibfield  {title} {\enquote {\bibinfo {title}
  {Laser-pulse-shape control of seeded {{QED}} cascades},}\ }\href {\doibase
  10.1038/s41598-017-05891-z} {\bibfield  {journal} {\bibinfo  {journal}
  {Scientific Reports}\ }\textbf {\bibinfo {volume} {7}},\ \bibinfo {pages}
  {5694} (\bibinfo {year} {2017})}\BibitemShut {NoStop}%
\bibitem [{\citenamefont {Poder}\ \emph {et~al.}(2018)\citenamefont {Poder},
  \citenamefont {Tamburini}, \citenamefont {Sarri}, \citenamefont {Di~Piazza},
  \citenamefont {Kuschel}, \citenamefont {Baird}, \citenamefont {Behm},
  \citenamefont {Bohlen}, \citenamefont {Cole}, \citenamefont {Corvan},
  \citenamefont {Duff}, \citenamefont {Gerstmayr}, \citenamefont {Keitel},
  \citenamefont {Krushelnick}, \citenamefont {Mangles}, \citenamefont
  {McKenna}, \citenamefont {Murphy}, \citenamefont {Najmudin}, \citenamefont
  {Ridgers}, \citenamefont {Samarin}, \citenamefont {Symes}, \citenamefont
  {Thomas}, \citenamefont {Warwick},\ and\ \citenamefont
  {Zepf}}]{poder_experimental_2018}%
  \BibitemOpen
  \bibfield  {author} {\bibinfo {author} {\bibfnamefont {K.}~\bibnamefont
  {Poder}}, \bibinfo {author} {\bibfnamefont {M.}~\bibnamefont {Tamburini}},
  \bibinfo {author} {\bibfnamefont {G.}~\bibnamefont {Sarri}}, \bibinfo
  {author} {\bibfnamefont {A.}~\bibnamefont {Di~Piazza}}, \bibinfo {author}
  {\bibfnamefont {S.}~\bibnamefont {Kuschel}}, \bibinfo {author} {\bibfnamefont
  {C.~D.}\ \bibnamefont {Baird}}, \bibinfo {author} {\bibfnamefont
  {K.}~\bibnamefont {Behm}}, \bibinfo {author} {\bibfnamefont {S.}~\bibnamefont
  {Bohlen}}, \bibinfo {author} {\bibfnamefont {J.~M.}\ \bibnamefont {Cole}},
  \bibinfo {author} {\bibfnamefont {D.~J.}\ \bibnamefont {Corvan}}, \bibinfo
  {author} {\bibfnamefont {M.}~\bibnamefont {Duff}}, \bibinfo {author}
  {\bibfnamefont {E.}~\bibnamefont {Gerstmayr}}, \bibinfo {author}
  {\bibfnamefont {C.~H.}\ \bibnamefont {Keitel}}, \bibinfo {author}
  {\bibfnamefont {K.}~\bibnamefont {Krushelnick}}, \bibinfo {author}
  {\bibfnamefont {S.~P.~D.}\ \bibnamefont {Mangles}}, \bibinfo {author}
  {\bibfnamefont {P.}~\bibnamefont {McKenna}}, \bibinfo {author} {\bibfnamefont
  {C.~D.}\ \bibnamefont {Murphy}}, \bibinfo {author} {\bibfnamefont
  {Z.}~\bibnamefont {Najmudin}}, \bibinfo {author} {\bibfnamefont {C.~P.}\
  \bibnamefont {Ridgers}}, \bibinfo {author} {\bibfnamefont {G.~M.}\
  \bibnamefont {Samarin}}, \bibinfo {author} {\bibfnamefont {D.~R.}\
  \bibnamefont {Symes}}, \bibinfo {author} {\bibfnamefont {A.~G.~R.}\
  \bibnamefont {Thomas}}, \bibinfo {author} {\bibfnamefont {J.}~\bibnamefont
  {Warwick}}, \ and\ \bibinfo {author} {\bibfnamefont {M.}~\bibnamefont
  {Zepf}},\ }\bibfield  {title} {\enquote {\bibinfo {title} {Experimental
  {{Signatures}} of the {{Quantum Nature}} of {{Radiation Reaction}} in the
  {{Field}} of an {{Ultraintense Laser}}},}\ }\href {\doibase
  10.1103/PhysRevX.8.031004} {\bibfield  {journal} {\bibinfo  {journal}
  {Physical Review X}\ }\textbf {\bibinfo {volume} {8}},\ \bibinfo {pages}
  {031004} (\bibinfo {year} {2018})}\BibitemShut {NoStop}%
\bibitem [{\citenamefont {Rackauckas}\ and\ \citenamefont
  {Nie}(2017)}]{rackauckas_differentialequationsjl_2017}%
  \BibitemOpen
  \bibfield  {author} {\bibinfo {author} {\bibfnamefont {C.}~\bibnamefont
  {Rackauckas}}\ and\ \bibinfo {author} {\bibfnamefont {Q.}~\bibnamefont
  {Nie}},\ }\bibfield  {title} {\enquote {\bibinfo {title}
  {{{DifferentialEquations}}.jl \textendash{} {{A Performant}} and
  {{Feature-Rich Ecosystem}} for {{Solving Differential Equations}} in
  {{Julia}}},}\ }\href {\doibase 10.5334/jors.151} {\bibfield  {journal}
  {\bibinfo  {journal} {Journal of Open Research Software}\ }\textbf {\bibinfo
  {volume} {5}},\ \bibinfo {pages} {15} (\bibinfo {year} {2017})}\BibitemShut
  {NoStop}%
\bibitem [{\citenamefont {Varin}\ and\ \citenamefont
  {Pich{\'e}}(2002)}]{varin_acceleration_2002}%
  \BibitemOpen
  \bibfield  {author} {\bibinfo {author} {\bibfnamefont {C.}~\bibnamefont
  {Varin}}\ and\ \bibinfo {author} {\bibfnamefont {M.}~\bibnamefont
  {Pich{\'e}}},\ }\bibfield  {title} {\enquote {\bibinfo {title} {Acceleration
  of ultra-relativistic electrons using high-intensity {{TM01}} laser beams},}\
  }\href {\doibase 10.1007/s00340-002-0906-8} {\bibfield  {journal} {\bibinfo
  {journal} {Applied Physics B}\ }\textbf {\bibinfo {volume} {74}},\ \bibinfo
  {pages} {s83--s88} (\bibinfo {year} {2002})}\BibitemShut {NoStop}%
\bibitem [{\citenamefont {Karmakar}\ and\ \citenamefont
  {Pukhov}(2007)}]{karmakar_collimated_2007}%
  \BibitemOpen
  \bibfield  {author} {\bibinfo {author} {\bibfnamefont {A.}~\bibnamefont
  {Karmakar}}\ and\ \bibinfo {author} {\bibfnamefont {A.}~\bibnamefont
  {Pukhov}},\ }\bibfield  {title} {\enquote {\bibinfo {title} {Collimated
  attosecond {{GeV}} electron bunches from ionization of high-{{Z}} material by
  radially polarized ultra-relativistic laser pulses},}\ }\href {\doibase
  10.1017/S0263034607000249} {\bibfield  {journal} {\bibinfo  {journal} {Laser
  and Particle Beams}\ }\textbf {\bibinfo {volume} {25}},\ \bibinfo {pages}
  {371--377} (\bibinfo {year} {2007})}\BibitemShut {NoStop}%
\bibitem [{\citenamefont {Fortin}, \citenamefont {Pich{\'e}},\ and\
  \citenamefont {Varin}(2009)}]{fortin_direct-field_2009}%
  \BibitemOpen
  \bibfield  {author} {\bibinfo {author} {\bibfnamefont {P.-L.}\ \bibnamefont
  {Fortin}}, \bibinfo {author} {\bibfnamefont {M.}~\bibnamefont {Pich{\'e}}}, \
  and\ \bibinfo {author} {\bibfnamefont {C.}~\bibnamefont {Varin}},\ }\bibfield
   {title} {\enquote {\bibinfo {title} {Direct-field electron acceleration with
  ultrafast radially polarized laser beams: Scaling laws and optimization},}\
  }\href {\doibase 10.1088/0953-4075/43/2/025401} {\bibfield  {journal}
  {\bibinfo  {journal} {Journal of Physics B: Atomic, Molecular and Optical
  Physics}\ }\textbf {\bibinfo {volume} {43}},\ \bibinfo {pages} {025401}
  (\bibinfo {year} {2009})}\BibitemShut {NoStop}%
\bibitem [{\citenamefont {Esarey}, \citenamefont {Schroeder},\ and\
  \citenamefont {Leemans}(2009)}]{esarey_physics_2009}%
  \BibitemOpen
  \bibfield  {author} {\bibinfo {author} {\bibfnamefont {E.}~\bibnamefont
  {Esarey}}, \bibinfo {author} {\bibfnamefont {C.~B.}\ \bibnamefont
  {Schroeder}}, \ and\ \bibinfo {author} {\bibfnamefont {W.~P.}\ \bibnamefont
  {Leemans}},\ }\bibfield  {title} {\enquote {\bibinfo {title} {Physics of
  laser-driven plasma-based electron accelerators},}\ }\href {\doibase
  10.1103/RevModPhys.81.1229} {\bibfield  {journal} {\bibinfo  {journal}
  {Reviews of Modern Physics}\ }\textbf {\bibinfo {volume} {81}},\ \bibinfo
  {pages} {1229--1285} (\bibinfo {year} {2009})}\BibitemShut {NoStop}%
\bibitem [{\citenamefont {Elkina}\ \emph {et~al.}(2011)\citenamefont {Elkina},
  \citenamefont {Fedotov}, \citenamefont {Kostyukov}, \citenamefont {Legkov},
  \citenamefont {Narozhny}, \citenamefont {Nerush},\ and\ \citenamefont
  {Ruhl}}]{elkina_qed_2011}%
  \BibitemOpen
  \bibfield  {author} {\bibinfo {author} {\bibfnamefont {N.~V.}\ \bibnamefont
  {Elkina}}, \bibinfo {author} {\bibfnamefont {A.~M.}\ \bibnamefont {Fedotov}},
  \bibinfo {author} {\bibfnamefont {I.~Y.}\ \bibnamefont {Kostyukov}}, \bibinfo
  {author} {\bibfnamefont {M.~V.}\ \bibnamefont {Legkov}}, \bibinfo {author}
  {\bibfnamefont {N.~B.}\ \bibnamefont {Narozhny}}, \bibinfo {author}
  {\bibfnamefont {E.~N.}\ \bibnamefont {Nerush}}, \ and\ \bibinfo {author}
  {\bibfnamefont {H.}~\bibnamefont {Ruhl}},\ }\bibfield  {title} {\enquote
  {\bibinfo {title} {{{QED}} cascades induced by circularly polarized laser
  fields},}\ }\href {\doibase 10.1103/PhysRevSTAB.14.054401} {\bibfield
  {journal} {\bibinfo  {journal} {Physical Review Special Topics - Accelerators
  and Beams}\ }\textbf {\bibinfo {volume} {14}},\ \bibinfo {pages} {054401}
  (\bibinfo {year} {2011})}\BibitemShut {NoStop}%
\bibitem [{\citenamefont {{Mercuri-Baron}}\ \emph {et~al.}(2021)\citenamefont
  {{Mercuri-Baron}}, \citenamefont {Grech}, \citenamefont {Niel}, \citenamefont
  {Grassi}, \citenamefont {Lobet}, \citenamefont {Piazza},\ and\ \citenamefont
  {Riconda}}]{mercuri-baron_impact_2021-1}%
  \BibitemOpen
  \bibfield  {author} {\bibinfo {author} {\bibfnamefont {A.}~\bibnamefont
  {{Mercuri-Baron}}}, \bibinfo {author} {\bibfnamefont {M.}~\bibnamefont
  {Grech}}, \bibinfo {author} {\bibfnamefont {F.}~\bibnamefont {Niel}},
  \bibinfo {author} {\bibfnamefont {A.}~\bibnamefont {Grassi}}, \bibinfo
  {author} {\bibfnamefont {M.}~\bibnamefont {Lobet}}, \bibinfo {author}
  {\bibfnamefont {A.~D.}\ \bibnamefont {Piazza}}, \ and\ \bibinfo {author}
  {\bibfnamefont {C.}~\bibnamefont {Riconda}},\ }\bibfield  {title} {\enquote
  {\bibinfo {title} {Impact of the laser spatio-temporal shape on
  {{Breit}}\textendash{{Wheeler}} pair production},}\ }\href {\doibase
  10.1088/1367-2630/ac1975} {\ \textbf {\bibinfo {volume} {23}},\ \bibinfo
  {pages} {085006} (\bibinfo {year} {2021})}\BibitemShut {NoStop}%
\bibitem [{\citenamefont {Arefiev}, \citenamefont {Robinson},\ and\
  \citenamefont {Khudik}(2015)}]{arefiev_novel_2015}%
  \BibitemOpen
  \bibfield  {author} {\bibinfo {author} {\bibfnamefont {A.~V.}\ \bibnamefont
  {Arefiev}}, \bibinfo {author} {\bibfnamefont {A.~P.~L.}\ \bibnamefont
  {Robinson}}, \ and\ \bibinfo {author} {\bibfnamefont {V.~N.}\ \bibnamefont
  {Khudik}},\ }\bibfield  {title} {\enquote {\bibinfo {title} {Novel aspects of
  direct laser acceleration of relativistic electrons},}\ }\href {\doibase
  10.1017/S0022377815000434} {\bibfield  {journal} {\bibinfo  {journal}
  {Journal of Plasma Physics}\ }\textbf {\bibinfo {volume} {81}} (\bibinfo
  {year} {2015}),\ 10.1017/S0022377815000434}\BibitemShut {NoStop}%
\bibitem [{\citenamefont {Tamburini}\ \emph {et~al.}(2010)\citenamefont
  {Tamburini}, \citenamefont {Pegoraro}, \citenamefont {Piazza}, \citenamefont
  {Keitel},\ and\ \citenamefont {Macchi}}]{tamburini_radiation_2010}%
  \BibitemOpen
  \bibfield  {author} {\bibinfo {author} {\bibfnamefont {M.}~\bibnamefont
  {Tamburini}}, \bibinfo {author} {\bibfnamefont {F.}~\bibnamefont {Pegoraro}},
  \bibinfo {author} {\bibfnamefont {A.~D.}\ \bibnamefont {Piazza}}, \bibinfo
  {author} {\bibfnamefont {C.~H.}\ \bibnamefont {Keitel}}, \ and\ \bibinfo
  {author} {\bibfnamefont {A.}~\bibnamefont {Macchi}},\ }\bibfield  {title}
  {\enquote {\bibinfo {title} {Radiation reaction effects on radiation pressure
  acceleration},}\ }\href {\doibase 10.1088/1367-2630/12/12/123005} {\bibfield
  {journal} {\bibinfo  {journal} {New Journal of Physics}\ }\textbf {\bibinfo
  {volume} {12}},\ \bibinfo {pages} {123005} (\bibinfo {year}
  {2010})}\BibitemShut {NoStop}%
\bibitem [{\citenamefont {Chen}\ \emph {et~al.}(2010)\citenamefont {Chen},
  \citenamefont {Pukhov}, \citenamefont {Yu},\ and\ \citenamefont
  {Sheng}}]{chen_radiation_2010}%
  \BibitemOpen
  \bibfield  {author} {\bibinfo {author} {\bibfnamefont {M.}~\bibnamefont
  {Chen}}, \bibinfo {author} {\bibfnamefont {A.}~\bibnamefont {Pukhov}},
  \bibinfo {author} {\bibfnamefont {T.-P.}\ \bibnamefont {Yu}}, \ and\ \bibinfo
  {author} {\bibfnamefont {Z.-M.}\ \bibnamefont {Sheng}},\ }\bibfield  {title}
  {\enquote {\bibinfo {title} {Radiation reaction effects on ion acceleration
  in laser foil interaction},}\ }\href {\doibase 10.1088/0741-3335/53/1/014004}
  {\bibfield  {journal} {\bibinfo  {journal} {Plasma Physics and Controlled
  Fusion}\ }\textbf {\bibinfo {volume} {53}},\ \bibinfo {pages} {014004}
  (\bibinfo {year} {2010})}\BibitemShut {NoStop}%
\bibitem [{\citenamefont {Mishra}, \citenamefont {Sharma},\ and\ \citenamefont
  {Sengupta}(2021)}]{mishra_effect_2021}%
  \BibitemOpen
  \bibfield  {author} {\bibinfo {author} {\bibfnamefont {S.~K.}\ \bibnamefont
  {Mishra}}, \bibinfo {author} {\bibfnamefont {S.}~\bibnamefont {Sharma}}, \
  and\ \bibinfo {author} {\bibfnamefont {S.}~\bibnamefont {Sengupta}},\
  }\bibfield  {title} {\enquote {\bibinfo {title} {Effect of radiation-reaction
  on charged particle dynamics in a focused electromagnetic wave},}\
  }\href@noop {} {\bibfield  {journal} {\bibinfo  {journal} {arXiv:2112.08063
  [physics]}\ } (\bibinfo {year} {2021})},\ \Eprint
  {http://arxiv.org/abs/2112.08063} {arXiv:2112.08063 [physics]} \BibitemShut
  {NoStop}%
\bibitem [{\citenamefont {Mishra}\ and\ \citenamefont
  {Sengupta}(2021)}]{mishra_exact_2021}%
  \BibitemOpen
  \bibfield  {author} {\bibinfo {author} {\bibfnamefont {S.~K.}\ \bibnamefont
  {Mishra}}\ and\ \bibinfo {author} {\bibfnamefont {S.}~\bibnamefont
  {Sengupta}},\ }\bibfield  {title} {\enquote {\bibinfo {title} {Exact solution
  of {{Hartemann}}\textendash{{Luhmann}} equation of motion for a charged
  particle interacting with an intense electromagnetic wave/pulse},}\ }\href
  {\doibase 10.1140/epjs/s11734-021-00260-4} {\bibfield  {journal} {\bibinfo
  {journal} {The European Physical Journal Special Topics}\ } (\bibinfo {year}
  {2021}),\ 10.1140/epjs/s11734-021-00260-4}\BibitemShut {NoStop}%
\bibitem [{\citenamefont {Yeh}\ \emph {et~al.}(2021)\citenamefont {Yeh},
  \citenamefont {Tangtartharakul}, \citenamefont {Rinderknecht}, \citenamefont
  {Willingale},\ and\ \citenamefont {Arefiev}}]{yeh_strong_2021}%
  \BibitemOpen
  \bibfield  {author} {\bibinfo {author} {\bibfnamefont {I.-L.}\ \bibnamefont
  {Yeh}}, \bibinfo {author} {\bibfnamefont {K.}~\bibnamefont
  {Tangtartharakul}}, \bibinfo {author} {\bibfnamefont {H.~G.}\ \bibnamefont
  {Rinderknecht}}, \bibinfo {author} {\bibfnamefont {L.}~\bibnamefont
  {Willingale}}, \ and\ \bibinfo {author} {\bibfnamefont {A.}~\bibnamefont
  {Arefiev}},\ }\bibfield  {title} {\enquote {\bibinfo {title} {Strong
  interplay between superluminosity and radiation friction during direct laser
  acceleration},}\ }\href {\doibase 10.1088/1367-2630/ac2394} {\bibfield
  {journal} {\bibinfo  {journal} {New Journal of Physics}\ }\textbf {\bibinfo
  {volume} {23}},\ \bibinfo {pages} {095010} (\bibinfo {year}
  {2021})}\BibitemShut {NoStop}%
\end{thebibliography}%
\end{document}